\documentclass[aps,prd,preprintnumbers,superscriptaddress,nofootinbib,floatfix,twocolumn,notitlepage,reprint]{revtex4-2}
\usepackage{graphicx,array}
\usepackage{bibunits}
\usepackage{hyperref}
\usepackage{xcolor}
\usepackage{amsmath,amssymb,slashed,latexsym}
\usepackage{bm}
\usepackage{braket}
\usepackage{placeins}
\usepackage{adjustbox}
\usepackage{makecell}
\usepackage[caption=false]{subfig}
\usepackage{enumitem}
\usepackage{amsthm}
\usepackage{comment}
\usepackage{multirow} 
\usepackage{tikz}
\usepackage[compat=1.1.0]{tikz-feynman}
\usepackage{soul}
\tikzstyle{block} = [draw, rectangle, 
minimum height=3em, minimum width=6em]

\newcommand{\inv}{{\rm inv}}
\newcommand*{\rom}[1]{\expandafter\@slowromancap\romannumeral #1@}

\newcommand{\cO}{\mathcal{O}}
\newcommand{\cL}{\mathcal{L}}

\newcommand{\fb}{\mathrm{fb}}

\newcommand{\BR}{\mathrm{BR}}

\newcommand{\MeV}{\mathrm{MeV}}
\newcommand{\GeV}{\mathrm{GeV}}

\newcommand{\ie}{\textit{i.e.}}

\def\beq{\begin{equation}}
\def\eeq{\end{equation}}
\def\beqa{\begin{eqnarray}}
\def\eeqa{\end{eqnarray}}

\makeatletter
\let\auto@bib@innerbib\@empty

\def\NAT@bibsetnum#1{%
  \setlength{\topsep}{\z@}%
  \NATx@bibsetnum{#1}%
}

\def\hyper@natlinkstart#1{%
  \Hy@backout{#1}%
  \hyper@linkstart{cite}{cite.\@bibunitname.#1}%
}
\def\hyper@natanchorstart#1{%
  \Hy@raisedlink{%
    \hyper@anchorstart{cite.\@bibunitname.#1}\hyper@anchorend}%
}

\newcommand{\putSMbib}{%
  \begingroup
    \let\SM@origlabel\label
    \def\label##1{%
      \def\SM@key{##1}%
      \def\SM@last{LastBibItem}%
      \ifx\SM@key\SM@last
        \SM@origlabel{SMLastBibItem}%
      \else
        \SM@origlabel{##1}%
      \fi
    }%
    \putbib
  \endgroup
}
\makeatother

\begin{document}

\preprint{CERN-TH-2025-238, MIT-CTP/5944}
\title{Braking protons at the EIC: from invisible meson decay to new physics searches}

\author{Reuven Balkin}
\affiliation{Department of Physics, University of California Santa Cruz and Santa Cruz Institute for Particle Physics, 1156 High St., Santa Cruz, CA 95064, USA}

\author{Ta'el Coren}
\affiliation{Physics Department, Technion - Israel Institute of Technology, Haifa 3200003, Israel}

\author{Alexander Jentsch}
\affiliation{Department of Physics, Brookhaven National Laboratory, Upton, NY 11973, USA}

\author{Hongkai Liu}
\affiliation{Department of Physics, Brookhaven National Laboratory, Upton, NY 11973, USA}

\author{Maksym Ovchynnikov}
\affiliation{Theoretical Physics Department, CERN, 1211 Geneva 23, Switzerland}

\author{Yotam Soreq}
\affiliation{Physics Department, Technion - Israel Institute of Technology, Haifa 3200003, Israel}
\affiliation{Theoretical Physics Department, CERN, 1211 Geneva 23, Switzerland}

\author{Sokratis Trifinopoulos}
\affiliation{Theoretical Physics Department, CERN, 1211 Geneva 23, Switzerland}
\affiliation{Center for Theoretical Physics -- a Leinweber Institute, Massachusetts Institute of Technology, Cambridge, MA 02139, USA}
\affiliation{Physik-Institut, Universit\"at Z\"urich, 8057 Z\"urich, Switzerland}

\begin{abstract}
We investigate the sensitivity of the Electron-Ion Collider~(EIC) to invisible final states in coherent exclusive electroproduction. 
The characteristic signal is a forward proton with reduced energy and little additional detector activity. 
Using the excellent particle detection capabilities and kinematics reconstruction at the EIC, we argue that backgrounds can be strongly suppressed.
While our analysis applies to various states, we specifically focus on vector and pseudoscalar particles: 
(i)~neutral mesons ($P = \pi^0,\eta^{(\prime)},
V = \rho^{0},\omega,\phi$), whose invisible Standard Model decays are extremely suppressed, and 
(ii)~gluon-coupled axion-like particles~(ALPs) decaying invisibly to a dark sector. 
With the baseline detector and a conservative residual-background estimate, the EIC can improve the sensitivity to invisible $\eta$, $\eta'$, $\omega$, and $\phi$ decays by factors of about $4$--$40$ and probe invisible $\rho^0$ decays for the first time. With further background reduction and optimized detector performance, the projected sensitivity may reach branching fractions as small as $4\times10^{-11}$.
In addition, under the same conditions, the EIC would directly probe invisibly decaying ALPs with the couplings up to $f_a\sim 10^5\,\GeV$ and masses in the range $0.1$--$2\,\GeV$.
\end{abstract}
\maketitle

\par{\textbf{Introduction.}} 
The upcoming Electron-Ion Collider~(EIC)~\cite{Accardi:2012qut,AbdulKhalek:2021gbh}, to be built at Brookhaven National Laboratory, will collide electrons of up to $18\,\GeV$ with protons of up to $275\,\GeV$ (or nuclei from deuterons to Uranium), eventually reaching annual integrated luminosities $\cL_{\rm int}=100\,\fb^{-1}$. 
Its physics program is primarily devoted to unraveling the partonic structure of nucleons and nuclei, but its kinematic reach and detector capabilities also make it a powerful laboratory for searches for rare and exotic processes, offering sensitive probes of physics beyond the Standard Model~(SM)~\cite{Gonderinger:2010yn,Boughezal:2020uwq,Liu:2021lan,Cirigliano:2021img,Davoudiasl:2021mjy,Yan:2021htf,Li:2021uww,Batell:2022ogj,Zhang:2022zuz,Yan:2022npz,Boughezal:2022pmb,Davoudiasl:2023pkq,Balkin:2023gya,Davoudiasl:2024vje,Wang:2024zns,Wen:2024cfu,Gao:2024rgl,Du:2024sjt,Deng:2025hio,Davoudiasl:2025rpn,Bellafronte:2025ubi,Jiang:2025frv,Huang:2025ljp,Bar-Shalom:2026fou,Adhikary:2026rck}.  

A distinctive feature of the EIC relevant for this work is its far-forward detector~(FFD) system~\cite{Jentsch:2023krakow,Pitt:2024utg}, which covers the forward-going protons ($\eta_{p'}>4.5$) up to very large pseudorapidities. 
This provides capabilities for exploring various signatures involving such protons. Coherent exclusive electroproduction with the \emph{invisible} final state $X$, 
\begin{align}
    \label{eq:process}
    p(p_p) + e^-(p_e) 
    \to 
    p(p_{p'}) + e^-(p_{e'}) + X(p_X)\, ,
\end{align}
stands out as an especially promising target. 

We consider two cases for the particle $X$.
\emph{Firstly}, we take it to denote the SM neutral mesons $\{\pi^0,\,\eta,\,\eta',\,\rho^{0}, \, \omega, \, \phi, \dots\}$.  
Such mesons are short-lived and decay visibly within the SM~\cite{Gao:2018seg}. 
However, invisible decay channels arise naturally in many extensions involving hidden sectors ~\cite{Fayet:2006sp,Kamenik:2011vy,Gninenko:2014sxa,Gninenko:2015mea,Gninenko:2016rjm,Barducci:2018rlx,Hostert:2020gou,REDTOP:2022slw}, including scenarios with decays to sterile neutrinos~\cite{Li:2020lba} or to particles that can accommodate the dark matter relic density~\cite{McElrath:2005bp,Batell:2014yra,Batell:2018fqo,Darme:2020ral,Ema:2020ulo}. 
\emph{Secondly}, we identify $X$ with a new, beyond-the-SM particle that can be copiously produced in $e\,p$ collisions and evades detection as well.  
For example, it could be a new spin-0 or spin-1 particle, which is coupled to hadrons and is either long-lived or decays into a dark sector.

\begin{figure}[t!]
    \centering
    \includegraphics[width=\linewidth]{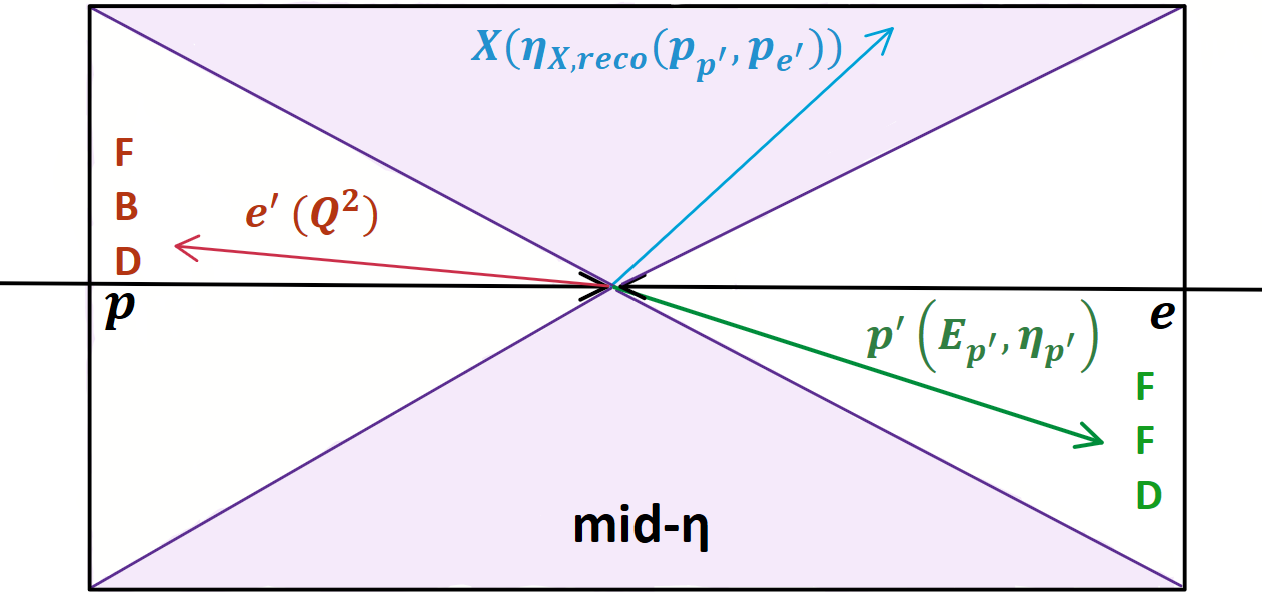}
    \caption{An illustration of the EIC missing-proton-energy~(MPE) search in $e+p \to e' +p' +X$ events. Without relying on reconstructing $X$, its pseudorapidity $\eta_{X}$ is indirectly extracted ($\eta_{X,{\rm reco}}(p_{e'},p_{p'})$) from the kinematics of outgoing states $p'$ and  $e'$. The latter are reconstructed by the far-forward (FFD) and central (mid-$\eta$)/far-backward (FBD) detectors if the event's $Q^{2},E_{p'},\eta_{p'}$ fall under detectability ranges (see Tbl.~\ref{tab:selection}).
    Events with $|\eta_X|<4$ are within the \emph{fully} detector-instrumented region; thus, visible $X$ can be vetoed, leaving only invisible states. See text for details. }
    \label{fig:scheme}
\end{figure}

If such invisible channels exist, the process in Eq.~\eqref{eq:process} would lead to a significant loss of proton energy with no accompanying detector activity. 
Current limits on invisible branching ratios of SM mesons~\cite{NA62:2020pwi,NA64:2024mah} leave ample room for the EIC to achieve major improvements.
In this work, we propose a \emph{missing-proton-energy}~(MPE) search strategy at the EIC, see Fig.~\ref{fig:scheme}, based on the efficient tagging of the forward proton by the FFD and on reconstructing the full event kinematics by additionally seeing the scattered electron. This provides significant background suppression and enables broad parameter space exploration. The uniqueness of the MPE over other missing energy searches at the EIC, such as the missing-electron-energy search in Ref.~\cite{Davoudiasl:2025rpn}, is its sensitivity to $X$ particles with hadronic interactions. This has important consequences both for the rare invisible meson decays and BSM searches.

Regarding $X$, we focus for concreteness on the pseudoscalar case (see Fig.~\ref{fig:diagram}), with mesons $P=\{\pi^0\,,\eta\,,\eta'\}$ and the axion-like particles~(ALPs) $a$ that couple to the SM primarily via gluons, see Eq.~\eqref{eq:Lagr_alp} below. In addition, we also provide our results for the case of a vector meson $V=\{\rho^0\,,\omega\,,\phi\}$.

\begin{figure*}[t!]
\centering
    \makebox[\textwidth][c]{
    \begin{minipage}[b]{0.55\textwidth}
        \centering
        \includegraphics[width=0.9\textwidth]{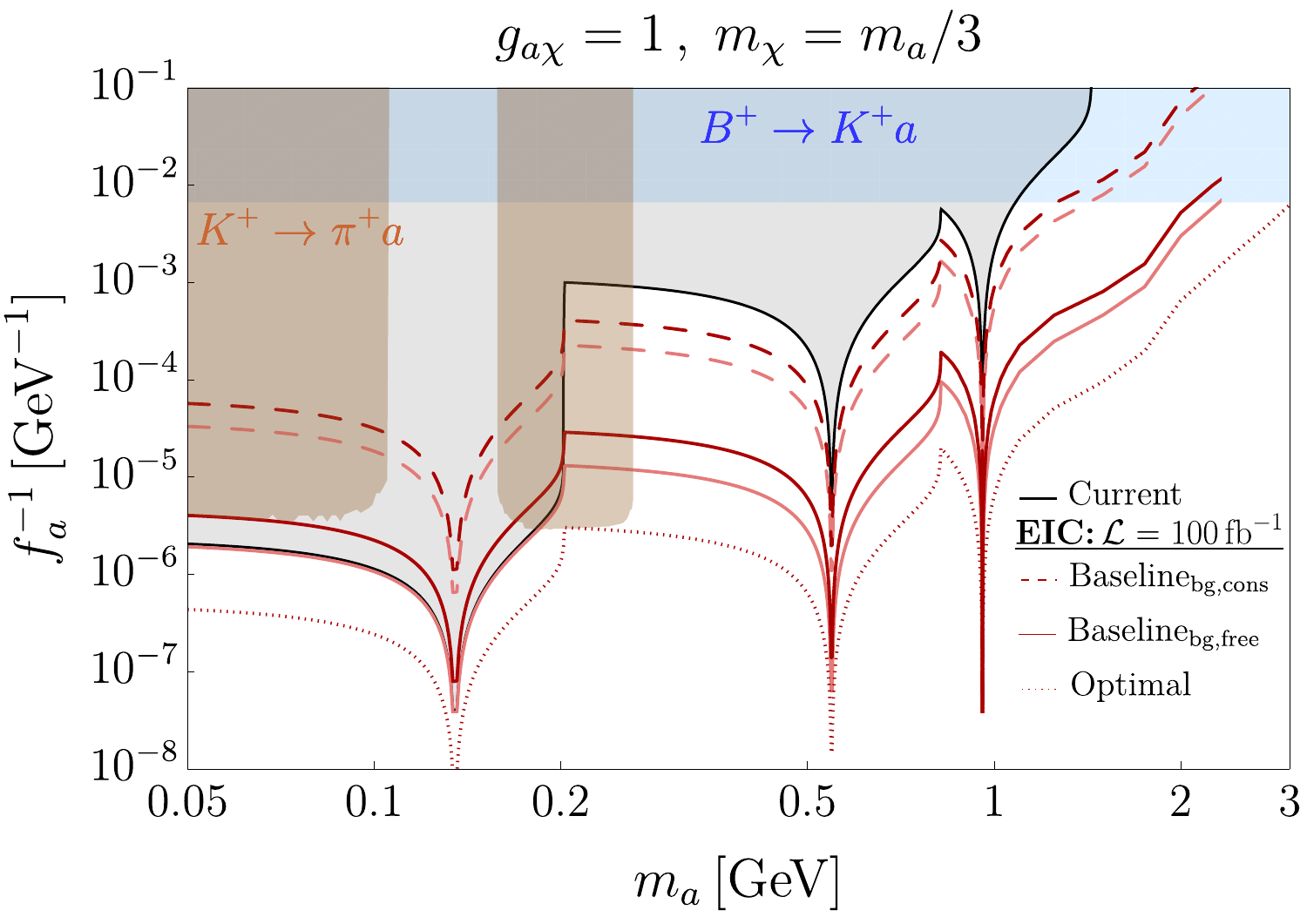}
    \end{minipage} \hspace{-1cm}
    \begin{minipage}[b]{0.55\textwidth}
        \centering
        \includegraphics[width=0.9\textwidth]{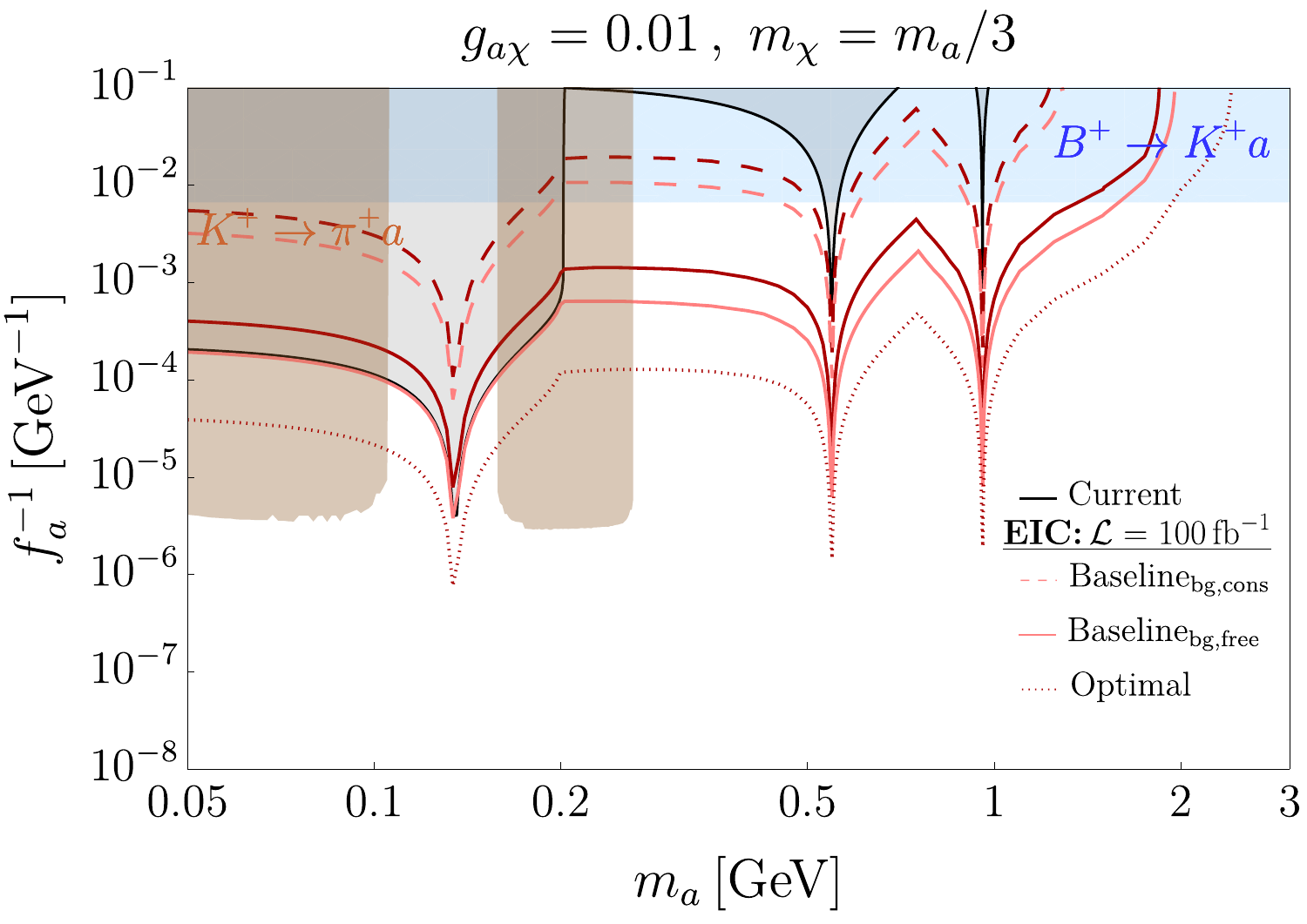}
    \end{minipage} }
	\caption{The EIC sensitivity to the model of invisibly decaying ALPs (see Eq.~\eqref{eq:Lagr_alp}) in the $\{m_a,f_a^{-1}\}$ plane, for $g_{a\chi}=1\,(0.01)$ on the left\,(right) panel. 
    The left\,(right) panel is dominated by invisible meson\,(ALP) decay. 
    We use the Baseline selection under conservative (Eq.~\eqref{eq:bg-conservative}) and background-free assumptions (dash and solid red), as well as the Optimal selection (red dotted).  
    Current bounds from invisible flavorless meson~\cite{NA62:2020pwi,NA64:2024mah} (black), kaon~\cite{NA62:2025upx} (brown region) and $B$-meson decays~\cite{Belle-II:2023esi} (turquoise region) are also presented. For the Baseline setup, the darker curves assume a setup from Tbl.~\ref{tab:selection}, while the lighter curves assume an upgraded FBD detector.}
	\label{fig:bound}
\end{figure*}

Our calculation of the coherent electroproduction cross sections for pseudoscalar states is based on the framework developed in the companion work~\cite{Balkin:2026whv} and can also be validated by a data-driven method. Our results, summarized in Tbl.~\ref{tab:mesons} and Fig.~\ref{fig:bound}, show that the same experimental analysis can improve the sensitivity to invisible $\eta$, $\eta'$, $\omega$, and $\phi$ decays by factors of about $4$--$40$ with the baseline detector and conservative background estimate, with substantially stronger reach under optimized benchmark assumptions, while also probing previously unexplored ALP parameter space. Supplemental Material~(SuM) provides details on the kinematics of the signal processes, as well as on simulations and background estimates.
\begin{figure}[t!]
\centering
\begin{minipage}[t]{0.98\columnwidth}
\centering
\begin{tikzpicture}[scale=0.75, every node/.style={transform shape}, baseline=(base)]
\begin{feynman}
  \coordinate (base) at (0,-1);
  \vertex (pL) at (-2,  1) {$p(p_p)$};
  \vertex (eL) at (-2, -1) {$e(p_e)$};
  \vertex (pR) at ( 2,  1) {$p(p_{p'})$};
  \vertex (eR) at ( 2, -1) {$e(p_{e'})$};
  \vertex (Vp) at (0,  1);
  \vertex (Ve) at (0, -1);
  \vertex (F)  at (0.5, 0);
  \vertex (X)  at (2, 0) {$X(k_X)$};
  \diagram*{
    (pL) -- [fermion] (Vp) -- [fermion] (pR),
    (eL) -- [fermion] (Ve) -- [fermion] (eR),
    (Ve) -- [boson, edge label={$\gamma$}] (F),
    (Vp) -- [boson, edge label'={$V$}] (F),
    (F)  -- [scalar] (X)
  };
\end{feynman}
\end{tikzpicture}
\end{minipage}
    \caption{
Feynman diagram for the electroproduction $e\,p \to e\,p X$ with $X=\pi^0,\,\eta^{(\prime)},\,a$ in the $V\gamma$-fusion channel, with $V=\rho^{0},\omega,\phi$, see Ref.~\cite{Balkin:2026whv} and text for details.}
    \label{fig:diagram}
\end{figure}

\textbf{Detector description.} 
The first experiment being constructed at the EIC is the Electron-Proton/Ion Collider~(ePIC) detector.
It contains elements of high-resolution tracking, electromagnetic and hadronic calorimetry, Cherenkov and time-of-flight particle identification, and beam-line detectors for tagging exclusive processes~\cite{ePICDetector}.
The EIC environment is effectively pileup-free, enabling straightforward timing- and vertex-based object reconstruction in the main detector.
The EIC may operate with a broad range of the incoming proton and electron energies $E_{p} = 41, 100-275\,\GeV$, $E_{e} = 5-18\,\GeV$. 
In addition to ePIC, there is a growing interest in building a second detector at the EIC.
It would be complementary to ePIC for discovery confirmation between two independent experiments, but with potentially different technologies to additionally focus on other measurements~\cite{BURKERT2023104032, NADEL-TURONSKI:2024/c}.

Considering the scattering process in Eq.~\eqref{eq:process}, we use the electron transferred momentum $Q^2\equiv-(p_e-p_{e'})^2$ and pseudorapidity $\eta_{e'}$, as well as the outgoing proton's energy $E_{p'}$ and pseudorapidity $\eta_{p'}$ to characterize the measured electron $e'$ and proton $p'$. Crucially, the exclusive electroproduction of hadronically interacting particles prefers $E_{e'}\approx E_{e}$~\cite{Balkin:2026whv}. 

In ePIC, the $e'$ may be reconstructed by one of the two detectors: by the central detector, or by the Far-Backward Detector (FBD). The coverage of the central detector is given by $|\eta_{e'}|<4$, while the FBD coverage is often defined in terms $10^{-3}< Q^2(E_{e'},\eta_{e'})/(1\,\GeV^2) < 0.1$. However, the FBD range does not accurately characterize the reconstruction efficiency for the particular kinematics $E_{e'}\approx E_{e}$, which is the case for the process in Eq.~\eqref{eq:process}. There, it is additionally limited by $-7.5<\eta_{e'}<-5.5$, and even inside this domain, the $e'$ particles have a small reconstruction efficiency~\cite{AbdulKhalek:2021gbh,Klest:2025fwx}. This is crucial not only for MPE but also for studying any coherent exclusive hadroproduction events at the EIC, including the Standard Model channels.

However, this limitation may be mitigated by optimized FBD instrumentation. 
To illustrate its potential, we examine the potential reach of an "upgraded FBD" in addition to the current instrumentation, under which we assume unit electron reconstruction efficiency in the window $Q^{2}>10^{-3}\,\GeV^{2}$, $\eta_{e'}<-4.2$.

On the proton side, the FFD system covers the proton-going direction to large pseudorapidities, with dedicated off-momentum proton tagging. 
In the current ePIC layout, protons with a momentum greater than 40\% of the beam momentum can be reconstructed with $\simeq 100\%$ efficiency for $\eta_{p'} > 4.5$ in the instrumented sections. The latter region could be extended with the second interaction region proposed in Ref.~\cite{PhysRevD.111.072013}. Besides the recoil proton, all other visible particles are reliably reconstructed by the mid-$\eta$ detector for $|\eta|<4$. 

Reconstructing $e',p'$, it is possible to indirectly reconstruct the kinematics of $X$ or its decay products even if they are not detected (which covers both the signal and background cases); in particular, the reconstructed pseudorapidity $\eta_{X,\text{reco}}(p_{e'},p_{p'})$ reproduces the true $\eta_{X}$ with a relative resolution of about $5\%$.

\begin{table}[t!]
\centering
\begin{tabular}{|c|c|c|}
  \hline
  Selection & \textbf{Baseline} & \textbf{Optimal} \\ \hline \hline
  $E_{p}, E_{e}$ & $(100, 10)$ GeV & $(41, 5)$ GeV \\ \hline
  $p', X$ & \multicolumn{2}{c|}{$\eta_{p'}>4.5,\; 0.5< E_{p'}/E_{p} <0.9,\; |\eta_{X}|<4$} \\ \hline
  $e'$  & \makecell{$|\eta_{e'}|<4$} & --\\ \hline
  Bkg  & $0$ (bg,free), Eq.~\eqref{eq:bg-conservative} (bg,cons) & 0\\ \hline
\end{tabular}
\caption{Two kinematics selection criteria of the outgoing proton $p'$, electron $e'$, and $X$ from the process of Eq.~\eqref{eq:process}, and backgrounds assumptions. For the Baseline setup, in Fig.~\ref{fig:bound}, we also include the sensitivity assuming the upgraded FBD setup, allowing to reconstruct $e'$ in the domain $\eta_{e'}<-4.2, \quad Q^{2} > 10^{-3}\,\text{GeV}^{2}$.}

\label{tab:selection}
\end{table}

\textbf{Background categories and proposed selection.}
SM channels with neutrinos in the final state, for example, decays of intermediate resonances into $\nu\bar\nu$, give an \emph{irreducible} contribution to the MPE events. 
Their cross sections are suppressed by the Fermi coupling and by phase space, and the expected yield over a full EIC run is well below one event; quantitative estimates are given in the SuM. 
The dominant concern is therefore \emph{reducible} background from processes that may mimic an invisible signature.

We group these backgrounds into two main classes: 
\begin{enumerate}
\item \emph{Beam-related and instrumental:} 
accidental overlaps (e.g., beam-gas collisions), particles traversing small uninstrumented gaps in fiducial acceptance, or rare reconstruction failures.
\item \emph{Leakage of visible states}: 
events of the type Eq.~\eqref{eq:process} in which the state $X$ is visible but not detected. 
This can occur either when $X$ (or its decay products) is produced at large pseudorapidity, $\eta_X \gtrsim 4$, where the detector coverage is incomplete as described above, or when $X$ lies in the fully instrumented region $|\eta_X|<4$ but its decay products fail to be reconstructed because of finite detection efficiency.
\end{enumerate}

A reliable determination of the first class of background requires detector runs; here we only outline general handles such as tight timing and vertex matching, global vetoes on additional activity in central and forward detectors, and the exclusion of events with elastic-like kinematics, for example $|\mathbf p_{T,e'}| \simeq |\mathbf p_{T,p'}|$. 
Further discussion is provided in the SuM.

Regarding the second class, we construct a selection to control the leakage of visible states without relying on any specific nature for $X$ (see Tbl.~\ref{tab:selection}). 
The key ingredients are: 
(i)~reconstruction of $e'$ and $p'$ with good energy and angular resolution; in what follows, we assume unit reconstruction efficiency for $e'$ and $p'$ within the specified domain;
(ii)~requiring $|\eta_{X,\text{reco}}|<4$ for the indirectly reconstructed pseudorapidity of the missing final state, for the event to be in the instrumented region; 
and (iii)~a veto on any additional reconstructed activity besides $e'$ and $p'$.

Two distinct selections are summarized in Tbl.~\ref{tab:selection}. 
The \emph{Baseline} choice corresponds to an early EIC running configuration with $(E_p,E_e)=(100,10)\,\GeV$ and considering the $e'$ detection solely by the central detectors. 
Additionally, to illustrate the impact of upgrading the FBD, we consider the Baseline selection with the upgraded FBD added, Baseline$_{\rm FBD}$. 
The \emph{Optimal} selection uses lower beam energies $(E_p,E_e)=(41,5)\,\GeV$ and assumes unit $e'$ reconstruction efficiency independently of its kinematics, to illustrate the maximal possible reach of the EIC, see the SuM.

Going forward, the selection efficiency could be further optimized using modern machine-learning techniques~\cite{Allaire:2023fgp,Araz:2024bom}.

\textbf{Background suppression.} 
To quantify the residual background after applying our selections, we generated samples of representative processes with visible final states $X$ (such as $\pi^0\to\gamma\gamma$, $\rho^0\to\pi^+\pi^-$, $\phi\to K\bar K$, and $\gamma^*\to\ell^+\ell^-$) and propagated them through a fast simulation of the ePIC detector response using the Baseline energy configuration $(E_p,E_e)=(100,10)\,\GeV$. The fast simulation includes beam energy spread, angular divergence, transport through the beam line, and the currently understood acceptances and resolutions. 
It has been benchmarked against full-simulation studies performed for other physics channels~\cite{PhysRevC.104.065205,dvcs2025}.

The quoted estimate concerns leakage of the simulated visible final states and is conservative for this background class: we use relatively high veto thresholds while machine-background studies are ongoing, and the reconstruction does not yet combine information from all ePIC subsystems.
Machine-induced and accidental backgrounds are a separate category that requires detector-specific studies with finalized geometry, digitization, and reconstruction.
Optimized thresholds and combined reconstruction, tighter fiducial cuts, and additional forward instrumentation are expected to reduce visible-state leakage further; details are given in the SuM.

In the simulation, we classify events according to the \emph{true} pseudorapidity of $X$: either $|\eta_X|<4$, where the detector is fully instrumented, or $\eta_X>4$, where only partial or no coverage is available. 
\emph{For events with $|\eta_X|<4$}, the rejection is limited by the efficiency of the global veto in the instrumented region. 
For the dominant $\pi^{0}\to \gamma\gamma$ background, full vetoing requires an inefficiency $\lesssim 10^{-9}$. Requiring that at least one photon be absorbed in the electromagnetic calorimetry leads to an effective depth of order ten radiation lengths, which is satisfied in ePIC for $|\eta_X|<4$.

We find in simulated events that decay products rarely escape without producing detectable signals in the tracking and/or calorimeter systems, for both non-muonic and muonic final states, where the latter rely on deposits in the hadronic calorimeters with $>4$ hadronic interaction lengths. 
Depending on the process, the probability of missing all visible activity in this region is $10^{-5}$--$10^{-3}$, mostly driven by near-edge events with $\eta_X\simeq 4$.
 
The corresponding conservative estimate of the background yield is
\begin{align}
   N_{\text{bg,cons}} \sim  9\times 10^{4}\,,
    \label{eq:bg-conservative}
\end{align}
mainly dominated by $\rho^{0}\to \pi^{+}\pi^{-}$, $\gamma^{*}\to \mu^{+}\mu^{-}$, and bremsstrahlung producing a photon. 
Further details are provided in the SuM.

\emph{For events with $\eta_X>4$}, contamination of the signal region requires both that all visible final states escape detection in uninstrumented or partially instrumented regions and that the event be misreconstructed as having $|\eta_{X,\text{reco}}|<4$. 
The latter probability is already small and can be further reduced by tightening the $\eta_{X,\text{reco}}$ requirement or improving the indirect reconstruction of $X$. 
We therefore assume that the dominant background comes from the $|\eta_X|<4$ region, with the estimate given in Eq.~\eqref{eq:bg-conservative}.

Given the discussion, we present projected sensitivities considering background assumptions: 
the conservative estimate Eq.~\eqref{eq:bg-conservative}, and the background-free hypothesis $N_{\text{bg}}=0$. 
To estimate the reach, we require $\sum_{X}N_{X}\times \BR(X\to\inv)>2\sqrt{N_{\text{bg}}}$ for the former scenario, and $\sum_{X}N_{X}\times \BR(X\to\inv)>3$ for the latter, where $N_{X} = \cL\times \sigma_{\rm sel}$, with $\sigma_{\rm sel}$ being the production cross section of the $X$ particle after imposing the selection from Tbl.~\ref{tab:selection}.


\begin{table}[t!]
\centering
\begin{tabular}{|c|c|c|}
\hline
$M$
& $\mathrm{BR}(M\to \mathrm{inv})_{\rm current}$
& $\mathrm{BR}(M\to \mathrm{inv})_{\rm EIC}$ \\
\hline\hline
$\pi^0$
& $4.4\times10^{-9}$~\cite{NA62:2020pwi}
& \makecell{
-- (Baseline$_{\text{bg,cons}}$)\\
-- (Baseline$_{\text{bg,free}}$)\\
$2\times10^{-10}$ (Optimal)
} \\
\hline
$\eta$
& $1.1\times10^{-4}$~\cite{NA64:2024mah}
& \makecell{
$2\times10^{-5}$ (Baseline$_{\text{bg,cons}}$)\\
$9\times10^{-8}$ (Baseline$_{\text{bg,free}}$)\\
$1\times10^{-9}$ (Optimal)
} \\
\hline
$\eta'$
& $2.1\times10^{-4}$~\cite{NA64:2024mah}
& \makecell{
$5\times10^{-5}$ (Baseline$_{\text{bg,cons}}$)\\
$2\times10^{-7}$ (Baseline$_{\text{bg,free}}$)\\
$4\times10^{-9}$ (Optimal)
} \\
\hline
$\rho^{0}$
& no bound
& \makecell{
$4\times10^{-7}$ (Baseline$_{\text{bg,cons}}$)\\
$2\times10^{-9}$ (Baseline$_{\text{bg,free}}$)\\
$4\times10^{-11}$ (Optimal)
} \\
\hline
$\omega$
& $7.3\times10^{-5}$~\cite{ParticleDataGroup:2024cfk}
& \makecell{
$2\times10^{-6}$ (Baseline$_{\text{bg,cons}}$)\\
$1\times10^{-8}$ (Baseline$_{\text{bg,free}}$)\\
$3\times10^{-10}$ (Optimal)
} \\
\hline
$\phi$
& $1.7\times10^{-4}$~\cite{ParticleDataGroup:2024cfk}
& \makecell{
$7\times10^{-6}$ (Baseline$_{\text{bg,cons}}$)\\
$4\times10^{-8}$ (Baseline$_{\text{bg,free}}$)\\
$3\times10^{-9}$ (Optimal)
} \\
\hline
\end{tabular}
\caption{Current experimental bounds on invisible decays of different mesons and the branching ratios the EIC can probe. We consider the Baseline and Optimal selections defined in Tbl.~\ref{tab:selection}; for Optimal, we assume zero background, $N_{\rm bg}=0$. A dash ``--'' in the last column indicates that the corresponding Baseline sensitivity does not improve the current bound.}
\label{tab:mesons}
\end{table}


\textbf{The pseudoscalar case.} 
Let us now study the implications of our analysis for the case of invisible decays of pseudoscalar states. 
These include invisible decays of $P = \pi^{0},\eta,\eta'$, as well as of hadronically-coupled ALPs $a$. 
We focus on the model of gluonic ALPs:
\begin{align}
    \label{eq:Lagr_alp}
	\mathcal L_a 
    \supset 
    \frac{\alpha_s}{4\pi } \frac{ a }{f_a} G^{\mu\nu} \tilde{G}_{ \mu \nu} 
    + ig_{a\chi}a\bar \chi \gamma^{5}\chi \,,
\end{align}
where ${\tilde{G}_{\mu\nu} \equiv (1/2) \epsilon_{\mu\nu \rho \sigma} G^{\rho \sigma}}$, $\chi$ is a dark sector particle, and the ALP mass $m_a$ is taken to be a free parameter. 
Motivated by the QCD axion~\cite{Peccei:1977hh,Peccei:1977ur,Weinberg:1977ma,Wilczek:1977pj} and its phenomenological implications~\cite{Abbott:1982af,Dine:1982ah,Preskill:1982cy,Marsh:2015xka,Adams:2022pbo,Nomura:2008ru,Freytsis:2010ne,Dolan:2014ska,Hochberg:2018rjs,Fitzpatrick:2023xks,Wang:2024mrc}, as well as by the axion quality problem~\cite{Kamionkowski:1992mf,Holman:1992us,Barr:1992qq,Ghigna:1992iv}, ALPs with masses much larger than the QCD axion $m_a^2 \gg  (m_\pi f_\pi/ f_a)^2$, with $f_a$ the ALP decay constant, have been considered. 
Such heavy variants arise in models addressing the strong CP problem~\cite{Dimopoulos:1979pp,Holdom:1982ex,Flynn:1987rs,Rubakov:1997vp,Berezhiani:2000gh,Fukuda:2015ana,Gherghetta:2016fhp,Dimopoulos:2016lvn,Fukuda:2017ywn,Agrawal:2017ksf,Gaillard:2018xgk,Lillard:2018fdt,Hook:2019qoh,Csaki:2019vte,Gherghetta:2020keg,Valenti:2022tsc,Kivel:2022emq,Dunsky:2023ucb}, and have been the focus of extensive theoretical and experimental studies~\cite{Dolan:2017osp,Alves:2017avw,Marciano:2016yhf,Jaeckel:2015jla,Dobrich:2015jyk,Izaguirre:2016dfi,Knapen:2016moh,Bauer:2018uxu,Mariotti:2017vtv,CidVidal:2018blh,Aloni:2018vki,Aloni:2019ruo,Bauer:2020jbp,Bauer:2021wjo,Sakaki:2020mqb,Florez:2021zoo,Brdar:2020dpr,Bertholet:2021hjl,Co:2022bqq,Trifinopoulos:2022tfx,Ghebretinsaea:2022djg,DallaValleGarcia:2023xhh,Kyselov:2025uez,Afik:2023mhj,Balkin:2021jdr,Blinov:2021say,Balkin:2023gya,Bai:2021gbm,Pybus:2023yex,Bai:2024lpq,Gao:2024rgl,Davoudiasl:2024fiz,Baruch:2025lbw,Ovchynnikov:2025gpx}. 
Moreover, we are interested in the regime where $\BR(a \to \chi\bar{\chi})\approx 1$ and the ALP does not decay visibly inside the detector, which is satisfied for most values of $g_{a\chi}$ due to the relatively small width to visible states $\Gamma_{a\to\text{vis.}}\ll m_a\sim  \Gamma_{a\to\bar{\chi} \chi}/g_{a\chi}^2$~\cite{Aloni:2018vki,Ovchynnikov:2025gpx,Balkin:2025enj}.

There are two contributions to MPE at the EIC in this model: 
decays of the on-shell ALP $X = a\to \text{inv}$ or of pseudoscalar mesons $X = P\to \text{inv}$, occurring via off-shell ALPs (\ie $\, P$-$a$ mixing), with the main production diagrams of $X$ particles shown in Fig.~\ref{fig:diagram}. 
Detailed cross-section calculations and kinematics analysis, used in this work, are performed in Ref.~\cite{Balkin:2026whv}. Namely, our calculation of the meson coherent electroproduction cross sections is based on a gauge-invariant extension of data-constrained photoproduction amplitudes~\cite{Yu:2016zut,Yu:2017vvp,Kashevarov:2017vyl,Wang:2025rvr} to the full $2\to3$ kinematics of Eq.~\eqref{eq:process}, appropriate for the low-$Q^2$ regime relevant for this search. We then construct the ALP production amplitude following the description of Refs.~\cite{Aloni:2018vki,Ovchynnikov:2025gpx,Balkin:2025enj}. 

The signal cross section, namely single-pseudoscalar production, can eventually be validated using meson decays into well-measured visible channels, thereby reducing the theoretical uncertainty on the signal yield. 
This approach is viable in our case thanks to the reliable reconstruction of the final-state system $X$ and the good invariant-mass resolution for all visible states. 
Notably, such a robust data-driven strategy can also facilitate precision measurements of rare Standard Model meson decays.

The EIC reach for invisible decays of pseudoscalar and vector mesons, assuming that only a single meson species contributes to the signal yield (which is often true for particular parameters of the model Eq.~\eqref{eq:Lagr_alp}), is summarized in Tbl.~\ref{tab:mesons}. Under the conservative background assumption and for the baseline ePIC setup, the EIC will be able to improve the reach of invisible decays of $\eta^{(')}$ ($\omega,\phi$) mesons by up to a factor of $\simeq 5$ ($20$--$40$), and, for the first time, explore invisible decays of $\rho^{0}$ mesons.
Reducing the background would improve the reach by a factor of about 200. Upgrading the FBD detector would add a factor of $2$--$5$, depending on the meson.
Relative to the Baseline$_{\rm FBD}$ benchmark, the Optimal selection would further improve the sensitivity by a factor of approximately $5$--$25$, depending on the meson, and would additionally allow a search for invisible $\pi^{0}$ decays.

For the $\pi^0$ channel, the Optimal setup probes somewhat lower $\sqrt{s_{\gamma^{*}p}}$ values, where our production framework discussed in Ref.~\cite{Balkin:2026whv} is less robust and subject to large uncertainties. 
However, this can be systematically studied by using visible $\pi^0$ decays of the same kinematics, which will be collected at the EIC.

The EIC probes the sum of the on-shell and off-shell ALP contributions in a complementary way, without the ability to disentangle them within a single energy run. 
The on-shell ALP signal yield scales as $\BR(a\to\text{inv})/f^2_a$, while the off-shell ALP yield scales as $g_{a\chi}^2/f_a^2$. 
Combining this with $\Gamma_{\text{vis.}}\ll \Gamma_{a\to\bar{\chi}\chi}/g^2_{a\chi}$, we find that the on-shell ALP production is the main probe for small values of $g_{a\chi}$, while the off-shell ALP dominates for $g_{a\chi}\sim\cO(1)$. 
In addition, the ALP mass reach of the two contributions is different as the off-shell production can probe masses above the mother meson at the $\cO(10\,\GeV)$ scale, see SuM. 

We present the combined sensitivity of the two contributions (on-shell and off-shell) on the $\{m_a,f_a^{-1}\}$ plane for $g_{a\chi}=1\,(0.01)$, as well as corresponding bounds, shown in the left\,(right) panel of Fig.~\ref{fig:bound}, see SuM for details.
In particular, the off-shell\,(on-shell) process dominates the sensitivity for $g_{a\chi}=1\,(10^{-2})$. 
Furthermore, we derive the bounds from meson invisible decays resulting from current experimental results on $K\to \pi+\rm{inv}$ from NA62~\cite{NA62:2025upx} and $\eta^{(\prime)}\to \text{inv}$ from NA64h~\cite{NA64:2024mah} (black region). 
Additionally, we show the current bounds inferred from the measurements $K^+ \to \pi^+ \nu \bar\nu$~\cite{NA62:2021zjw} (brown region) and $B^+ \to K^+ \nu \bar\nu$~\cite{Chakraborty:2021wda,Belle-II:2023esi,Altmannshofer:2023hkn,Abumusabh:2025zsr} (turquoise region).

If an MPE excess were observed, varying the beam energies could help distinguish signal contributions despite the limited missing-mass resolution. Over $\sqrt{s}=28$--$140\,\GeV$, both the event rate and the missing-energy distribution $E_X=E_p+E_e-E_{p'}-E_{e'}$ depend on the missing-particle mass; for example, under the Baseline selection the $\pi^0$ rate is about three times the $\eta'$ rate~\cite{Balkin:2026whv}. A dedicated analysis is left for future work.

\textbf{Outlook.}
We have introduced the missing-proton-energy strategy for invisible final states produced in coherent electroproduction at the EIC. Even with the baseline detector and conservative background assumptions, it substantially improves the sensitivity to invisible pseudoscalar- and vector-meson decays and probes new ALP parameter space. Such sensitivities usually require dedicated experiments, while no approved high-statistics program currently targets comparable reach across the $\eta$, $\eta'$, $\rho^0$, $\omega$, and $\phi$ channels. Production uncertainties can be reduced by normalizing to visible decays measured in the same topology.

Improved background rejection, optimized far-backward electron reconstruction, and a complementary second detector could extend the reach and help distinguish meson and ALP contributions~\cite{BURKERT2023104032,NADEL-TURONSKI:2024/c,Chang:2025pgi}. The same strategy applies to other hadronically coupled invisible states, including spin-0, spin-1, and millicharged particles, and can be combined with missing-electron-energy searches when both hadronic and leptonic interactions are present~\cite{Davoudiasl:2025rpn}. It may also be adapted to future lepton--hadron facilities with different energies or beam species~\cite{Anderle:2021wcy,LHeC:2020van,Acosta:2022ejc,Davoudiasl:2024fiz,Akturk:2025ubm}.

\begin{acknowledgments}
We are grateful to Admir Greljo, Or Hen, Tim Hobbs, Rotem Ovadia, Fernando Romero-López, Miguel Vanvlasselaer, Urs Wiedemann, and Mike Williams for their helpful comments and encouragement.
The research of RB is supported in part by the U.S. Department of Energy grant number DE-SC0010107. HL is supported by the U.S. Department of
Energy under Grant Contract DE-SC0012704. 
TC and YS are supported by the ISF (grant No. 597/24) and by BSF (grant No. 2024091).
YS thanks CERN-TH for the scientific associateship.
ST was supported by the Office of High Energy Physics of the U.S. Department of Energy (DOE) under Grant No.~DE-SC0012567, and by the DOE QuantISED program through the theory consortium “Intersections of QIS and Theoretical Particle Physics” at Fermilab (FNAL 20-17).
ST is additionally supported by the Swiss National Science Foundation - project n. P5R5PT\_222350, and acknowledges the CERN TH Department for hospitality while this research was being carried out. 
The work of AJ is supported by the U.S. Department of Energy under Contract DE-SC0012704.
This project has also received funding from the European Union’s Horizon Europe research and innovation programme under the Marie Skłodowska-Curie Staff Exchange grant agreement No 101086085 - ASYMMETRY.
\end{acknowledgments}
\bibliographystyle{JHEP}
\bibliography{main}

\clearpage

\appendix

\onecolumngrid
\section*{Supplemental material}

\setcounter{equation}{0}
\setcounter{figure}{0}
\setcounter{table}{0}
\setcounter{page}{1}
\renewcommand{\thepage}{S\arabic{page}}
\makeatletter
\renewcommand{\thefigure}{S\arabic{figure}}
\renewcommand{\thetable}{S\arabic{table}}
\renewcommand{\theHfigure}{S\arabic{figure}}
\renewcommand{\theHtable}{S\arabic{table}}
\makeatother

In this Supplemental Material, we provide technical details on the calculations of event rates and sensitivity of the EIC.
Sec.~\ref{app:backgrounds} is devoted to discussing the Standard Model~(SM) backgrounds for the missing-proton-energy~(MPE) search at the Electron-Ion Collider~(EIC). 
In Sec.~\ref{app:event-selection}, we discuss the impact of the Baseline selection (Tbl.~\ref{tab:selection} of the main text) on the event yield, highlighting in particular the effect of the $Q^{2}$ cut and the role of the beam energies in maximizing the signal-to-background ratio. 
Sec.~\ref{app:meson-to-invis} discusses details about the two contributions to the MPE signature at the EIC for the ALP model (ALP decays and pseudoscalar meson decays via off-shell ALPs). 
In particular, it presents the expression for the branching of invisible decays of pseudoscalar mesons via the ALPs with an arbitrary coupling pattern, and discusses the complementarity between the ALP and meson decays. 
Finally, in Sec.~\ref{app:na64h}, we estimate the bounds on the ALP model from ALP-on-shell and ALP-off-shell decays from NA62 and NA64.

\section{Further details on backgrounds}
\label{app:backgrounds}

In this section, we discuss various background sources for the MPE searches, including irreducible backgrounds, events from physical processes that may mimic the missing decays, and possible instrumental sources. 
The main purpose of the discussion is to demonstrate that the background can be well-controlled with the currently planned setup of the EIC detector.
We also discuss ways in which it can be improved, either via selection criteria or different experimental considerations.
Throughout this section, we consider the Baseline selection from Tbl.~\ref{tab:selection}, corresponding to the energy configuration with $E_{p} = 100\,\GeV$ and $E_{e} = 10\,\GeV$.

\subsection{Irreducible backgrounds}

An \emph{irreducible} background comes from the events of the type 
\begin{align}
    e p 
    \to 
    e p + \nu\bar{\nu}/2(\nu\bar{\nu}) \, .
\end{align}
Such events may be either direct (with the off-shell $Z$ boson), or involve an intermediate on-shell resonance $X = \{\pi^{0},\eta^{(\prime)},\rho^{0},\dots\}$, which then decays to neutrinos via its mixing with the $Z$ boson. 
Using naive dimensional analysis, we estimate $\sigma \sim G_{F}^{2}s_{\nu\nu}\alpha_{\text{EM}}^{2}/[24\cdot (4\pi)^{3}]$, with $s_{\nu\nu}<s_{ep}$ being the invariant mass of the collision. 
It gives us $<1$ event for $\cL_{\rm int}=100\,\fb^{-1}$. 
We confirm this by simulating the corresponding inelastic scattering process in MadGraph5\_aMC@NLO~\cite{Alwall:2014hca}, and obtaining the total cross section $\sigma(ep\to ep\nu\bar\nu)\sim{3\times10^{-4}\,\fb}$.
As for the decays via intermediate particles (e.g., $\rho^{0}\to \nu\bar{\nu}$), the invisible branching ratios of various mesons are at the level of $10^{-11}$ or smaller~\cite{Gao:2018seg}, which also implies $<1$ events per full EIC running time.
Therefore, we conclude that the irreducible background is negligible.

\subsection{Particle leakage} 

\textbf{Leakage sources and partial instrumentation.} 
Another class of background events consists of low-$Q^{2}$ processes with emission of the visible states that escape detection. 
It can be a result of an uninstrumented domain of the detector or of imperfections in the detection of activity in the instrumented region.
 
Let us first describe the instrumentation.
Besides the proton, charged-particle reconstruction may be reliable up to pseudorapidities $\eta \approx 6$. 
However, in the domain $4.5 \lesssim \eta \lesssim 6$, the FFD provides only partial azimuthal coverage $|\Delta\phi_{\rm FFD}|/2\pi \sim 0.5$. 
The higher-$\eta$ region is generally off-limits for other charged particles because they must pass through the hadron beam line to reach the FFDs. 
In practice, charged particles with mass different than the proton mass and/or with momentum less than about $40\%$ of the primary beam momentum are generally not seen in the FFDs, due to their different magnetic rigidity as they traverse the beam-line magnets.

Neutral-particle detection in the forward hemisphere is nearly hermetic thanks to the FFD calorimeters.  
In particular, forward photons achieve near-optimal efficiency for $\eta \gtrsim 6$. 
In the domain $4.5 \lesssim \eta \lesssim 6$, the same partial azimuthal coverage noted for charged particles applies, and in the window $4 < \eta < 4.5$ the coverage vanishes completely. 
At $|\eta| < 4$, particles may instead be reconstructed by the mid-$\eta$ detectors.

As discussed in the main text, such a background is reduced by the selection from Tbl.~\ref{tab:selection}. 
The selection suppresses particle leakage through uninstrumented regions, while the events in the instrumented regions may be vetoed if requiring zero activity behind the outgoing electron and proton, including displaced deposits from decays of long-lived strange states. 

The measurements' resolution is as follows~\cite{AbdulKhalek:2021gbh}. For the electron in the relevant $Q^2$ domain for the main ePIC detector, backward tracking and electromagnetic calorimeter~(ECAL) measure $E_{e'}$, $\theta_{e'}$, and $\phi_{e'}$ with precision $\delta E_{e'}/E_{e'} = (1\%)/\sqrt{E_{e'}} \oplus 1\% $ and ${\delta[\tan\theta_{e'}]}/{\tan\theta_{e'}}\sim2\%$ (similar for $\phi_{e'}$), where $\delta[x]$ denotes the combined uncertainty on $x$. In the FBD domain, the energy resolution worsens to $\delta E_{e'}/E_{e'} \sim 10\% $ and the angular resolution worsens to ${\delta[\tan\theta_{e'}]}/{\tan\theta_{e'}}\sim5\%$ (similar for $\phi_{e'}$).
The forward proton is reconstructed via transport through the machine optics to yield $(E_p',\eta_{p'})$ with ${\delta [p_{T, p'}]}/{p_{T, p'}}\sim5\%$ for a transverse momentum $p_{T, p'}\sim 1\,\GeV$ and ${\delta [p_{p'}]}/{p_{p'}}\sim3\%$.
Reconstructing $e',p'$, it is possible to indirectly reconstruct the kinematics of $X$ or its decay products even if they are not detected (which covers both the signal and background cases); in particular, the reconstructed pseudorapidity $\eta_{X,\text{reco}}(p_{e'},p_{p'})$ reproduces the true $\eta_{X}$ with a relative resolution of about $5\%$.
Finite resolution effects can easily be mitigated by taking a slightly tighter cut on $\eta_X$, which only has a small effect on the signal efficiency. 
Once left with events that truly lie within the instrumented region, the only remaining source of leakage (\ie{} vetoing inefficiency) is due to the particle detection efficiencies of the detector system, to be discussed below.

\textbf{Contributing processes and the vetoing inefficiency in $|\eta|<4$ region.} 
Given the range of detectable $Q^{2}$ we consider, and also the fact that the probability of having such fake-invisible events decreases with the final state multiplicity, the processes causing backgrounds are coherent exclusive photoproduction. They include
\begin{align}
    \label{eq:quasi-el-1}
    e + p &\to e + p + M, \quad M \to \text{vis.}\,, \\ 
    \label{eq:quasi-el-0}
    e + p &\to e + p + \gamma\,, \\
    \label{eq:quasi-el-pair}
    e + p &\to e + p + \gamma^{*} \to e + p + \ell^{+} + \ell^{-}\,, \\
    \label{eq:quasi-el-di-nucleon}
    e + p &\to e + p + N + \bar{N}\,,  
\end{align}
where $M$ is a meson decaying into any visible final state (e.g., $\pi^{0}\to \gamma\gamma$, $\eta \to \mu^{+}\mu^{-}$, $\rho^{0}\to \pi^{+}\pi^{-}$, $\phi \to K\bar{K}$, etc.) and $N$ is a nucleon. 
Qualitatively, these processes may be classified into those with neutral non-hadronic states, electron-positron and muon-antimuon pairs, charged light hadronic particles, states with strange mesons, and heavy hadronic states.

\begin{table}[t!]
    \centering
    \begin{tabular}{|c|c|c|}
     \hline Final state $X$  & $\sigma_{\text{sel}}$ [fb] &$\sigma_{\rm sel}\times \epsilon_{\rm ineff}$ [fb]  \\ \hline
    $\gamma$  & $1.5\times 10^{5}$ & $3\times 10^{2}$ \\ \hline
    $\gamma^{*}\to e^{+}e^{-}$  & $\sim 5\times 10^{3}$ & $\sim 3\times 10^{-2}$ \\ \hline
    $\gamma^{*}\to\mu^{+}\mu^{-}$ & $2\times 10^{4}$ & $50$ \\ \hline
    $\pi^{0}\to \gamma\gamma$  & $4\times 10^{6}$ & $10$ \\ \hline
   $\pi^{0}\to e^{+}e^{-}$  & $0.3$ & $1 \times 10^{-6}$ \\ \hline
    $\eta\to \mu^{+}\mu^{-}$  & $2$ & $6\times 10^{-3}$ \\ \hline
    $\rho^{0}\to \pi^{+}\pi^{-}$  & $2\times 10^{7}$ & $5\times 10^{2}$ \\ \hline
    $\phi \to K\bar{K}$  & $1\times 10^{6}$ & $15$ \\ \hline
    \end{tabular}
    \caption{Cross sections for various quasi-elastic processes after imposing the Baseline selection from Tbl.~\ref{tab:selection} (second column) and if additionally multiplying by the vetoing inefficiency $\epsilon_{\text{ineff}}$, see text for details. The number of events per full EIC run may be obtained by multiplying the cross sections by the total luminosity $\cL=100\,\fb^{-1}$. The sum of the cross sections from the third column, $\sigma_{\text{sel}}\times\epsilon_{\text{ineff}}$, multiplied by the total luminosity, gives the estimate Eq.~\eqref{eq:bg-conservative} from the main text.}
    \label{tab:quasi-elastic-bg}
\end{table}

To test the efficiency of the Baseline selection in the $\eta_{X}<4$ domain, we proceed as follows. 
First, we compute the rates of the processes Eqs.~\eqref{eq:quasi-el-0}--\eqref{eq:quasi-el-pair} by calculating their total cross sections and the fraction of events that survive the selection.

The cross sections for the meson-production processes Eq.~\eqref{eq:quasi-el-1} are obtained using the methodology of Ref.~\cite{Balkin:2026whv}. Namely, it calculates the meson coherent electroproduction cross sections based on a gauge-invariant extension of data-constrained photoproduction amplitudes~\cite{Yu:2016zut,Yu:2017vvp,Kashevarov:2017vyl,Wang:2025rvr} to the full $2\to3$ kinematics of the process from Eq.~\eqref{eq:process}, appropriate for the low-$Q^2$ regime relevant for this search.

We consider $\pi^{0}\to \gamma\gamma/e^{+}e^{-}$, $\eta\to \mu^{+}\mu^{-}$, as well as $\rho^{0}\to \pi^{+}\pi^{-}$ and $\phi \to K^{+}K^{-}/K_{L}K_{S}$. 
As for pair production via $\gamma^{*}$ and bremsstrahlung producing $\gamma$, Eqs.~\eqref{eq:quasi-el-0},~\eqref{eq:quasi-el-pair}, we generate the kinematics of the corresponding processes using \textsc{MadGraph5\_aMC@NLO}~\cite{Alwall:2014hca} with a \textsc{FeynRules} implementation~\cite{Alloul:2013bka,Christensen:2008py}, treating the proton as a point-like particle; this approximation is adequate in the relevant $Q^{2}$ range. 
Most of the total cross section of these processes (before imposing any cuts) is dominated by the very soft final state particles $\gamma, \ell^{+}\ell^{-}$, and the energy-based selection reduces it by a few orders of magnitude. 

Finally, processes of the type~\eqref{eq:quasi-el-di-nucleon} require a large invariant-mass transfer and are expected to be subdominant compared to the other channels; we therefore do not consider them in detail and list them only for completeness.

The resulting yields, reported in the second column of Tbl.~\ref{tab:quasi-elastic-bg}, correspond to the number of events after the Baseline selection but \emph{before any detector-activity veto}. Per full EIC run, $\cL = 100\text{ fb}^{-1}$, they are at the level of: $\simeq 10^{9}$ events for the neutral final states (set by the $\pi^{0}\to \gamma\gamma$ process), $\simeq 10^{9}$ events for the charged non-muonic final states (set by $\rho^{0}\to \pi^{+}\pi^{-}$), and $\simeq 10^{6}$ for the muonic final states (set by the $\gamma^{*}\to \mu^{+}\mu^{-}$ process).
 
Next, to evaluate the efficiency of vetoing these remaining events in the planned ePIC setup, we generate event samples for several processes with $\eta_{X}<4$ belonging to the instrumented regions, to process them with a parametrized fast simulation of ePIC, validated using the ePIC software framework~\cite{ref:ePIC_dd4hep}, similar to Ref.~\cite{dvcs2025}. 
The processing mimics detector-level and event reconstruction effects: 
it adds beam-related effects such as energy spread and angular divergence, particle transport, and finite reconstruction efficiencies.

The main caveats in this estimate concern the detector-activity thresholds and the use of a parametrized reconstruction rather than a unified algorithm that exploits the full set of ePIC subsystems.
The veto thresholds are chosen conservatively -- that is, at relatively high values -- because ePIC studies of machine-induced activity, including beam-gas collisions, are ongoing and the associated data rates have not yet fixed the lowest operational thresholds.
This choice reduces the simulated veto efficiency and therefore increases the estimated visible-state leakage.
Likewise, the present reconstruction does not yet use complementary information from particle-identification and timing detectors, such as time-of-flight and Cherenkov systems, which may veto an event even when activity is not reconstructed in the tracker or calorimeters.

For each process, we generate $N_{\text{sample}} = 10^{6}$ events using \textsc{MadGraph} and estimate the effect of the kinematic cuts by imposing all relevant pre-selections on lab-frame quantities (e.g.\ energies).
We can then use the resulting distributions to estimate channel-specific particle detection efficiencies as discussed above.
However, in practice, we use only the sample from the decaying-meson channel \eqref{eq:quasi-el-1}, which contains a sufficient number of events.
Due to the limitations of \textsc{MadGraph}, generating a comparable sample size for ~\eqref{eq:quasi-el-0} and~\eqref{eq:quasi-el-pair} would be computationally very expensive. 
Since we expect per-particle detection efficiencies to be only mildly different across different channels, we simply use the same detection efficiencies found for the decaying-meson channel \eqref{eq:quasi-el-1} for the remaining background channels~\eqref{eq:quasi-el-0} and~\eqref{eq:quasi-el-pair}.

For a simple veto required to suppress the background, only a single track or calorimeter energy deposit is sufficient to reject a background process. 
For charged particles with $\eta < 3.7$, the tracking detectors will be able to reconstruct a particle trajectory, with additional information from calorimeter energy deposits also present, and extending beyond the tracker acceptance to $\eta < 4.0$. 
This makes vetoing straightforward because multiple detector subsystems will record relevant information used for the veto. 
For photons, only energy deposits in the electromagnetic calorimeters will be useful for vetoing. However, given the large number of radiation lengths present, the efficiency is still expected to be high.

From the simulation, we adopt the following single particle detection efficiencies for particles in the fiducial acceptance (in order from highest to lowest efficiency, based on ePIC detector instrumentation in the fiducial single-particle acceptance): 
i) $\epsilon_{\gamma} ~\sim 99.8\%$ 
ii) $\epsilon_{e^{\pm}} \sim 99.8\%$, 
iii) $\epsilon_{h^{\pm}} \sim 99.5\%$ for charged hadrons $h$, 
and iv) $\epsilon_{\mu^{\pm}} \sim 95\%$. 
For non-muonic states, these numbers are much tighter than the ones considered in Ref.~\cite{dvcs2025}, where a 95\% single particle efficiency was assumed in the main ePIC detector volume for a fully reconstructed photon. 
Indeed, for our purposes, we do not require a full reconstruction of particle kinematics but just a simple veto. 
In the case of muons, we assume a more conservative efficiency given how many of the $\mu^{+}\mu^{-}$ events are near the edge of the ePIC acceptance region of $\eta < 4.0$. 
In cases where only one muon is within that acceptance, and near the edge, we place a $100\,\MeV$ threshold cut on its energy deposit to decide if that muon would be seen by ePIC. 
This leads to a loss of efficiency $\sim 1\%$ compared to a case where this threshold is not considered. 
The overall vetoing efficiency could be improved by tightening the meson production cut to be well within the ePIC main detector acceptance (e.g., from $\eta < 4.0$ to $\eta < 3.7-3.8$), ensuring all particles have access to tracking information, in addition to calorimetry energy deposits. Vetoing efficiencies for mesons produced within $\eta < 4.0$ are used in this study.

The resulting vetoing inefficiencies are $\epsilon^{\gamma}_{\text{ineff}} = 2\times 10^{-3}$ for bremsstrahlung producing a photon, $\epsilon^{\gamma\gamma}_{\text{ineff}}=\epsilon^{e^{+}e^{-}}_{\text{ineff}} = 4\times 10^{-6}$ for decays into two photons or into the electron-positron pair, $2.5\times 10^{-5}$ into charged hadrons, and $2.5\times 10^{-3}$ for the muons. Using the reported efficiencies, we present the residual background yields from the events~\eqref{eq:quasi-el-0}-\eqref{eq:quasi-el-di-nucleon} in Tbl.~\ref{tab:quasi-elastic-bg}. 

A future study with finalized ePIC geometry, digitization, detector thresholds, realistic machine backgrounds, and unified reconstruction will refine the total experimental background estimate.
For the visible-state leakage considered here, combining information across subsystems and lowering thresholds as permitted by machine conditions are expected to improve the rejection.
Additional gains can be obtained by tightening the $\eta_{X,\text{reco}}$ requirement to keep events further inside the fully instrumented region and by adding dedicated forward muon and photon detectors, either as a future ePIC upgrade or in a possible second EIC detector~\cite{NADEL-TURONSKI:2024/c}.
Thus, the reported visible-state leakage estimate is conservative for the simulated processes, while machine-induced and accidental backgrounds constitute a separate category discussed below.

\subsection{Instrumental and reconstruction effects} 
The last category is instrumental sources. 
They can mimic the exclusive missing-energy topology even in the absence of genuine invisible states. Below, we discuss the potential origins of such backgrounds and sketch the strategies that could mitigate them; a more quantitative discussion would require a full detector simulation, which is beyond the scope of this work.

Three categories are relevant:

\emph{(1) Accidental overlaps and beam-related activity.}  
Beam-gas and beam-halo interactions may produce in-time signals in forward proton detectors uncorrelated with the hard interaction.  
A coincident off-momentum proton can be falsely associated with a genuine $e+p\to e+X$ scattering lacking a leading proton, fabricating an exclusive $e+p\to e'+p'$ candidate with apparent missing energy. Likewise, elastic $e+p\to e'+p'$ events can be misclassified as missing energy events if unrelated neutral energy deposits (from beam losses or beam-gas interactions) compromise the forward veto logic. 

These classes are suppressed by requiring tight bunch-crossing timing for the final-state proton and electron; consistency of the measured $\xi = E_{p'}/E_{p}$ and $Q^2$ with proton transport from the reconstructed vertex; and the absence of any extra detector activity above low thresholds in central and forward calorimetry and tracking.

\emph{(2) Material interactions and local inefficiencies.}  
Photons emitted at small polar angles may convert upstream in windows or beam pipe components, with secondaries steered out of instrumented acceptance by magnetic elements.  
Tight collimation of photons from $P=\{\pi^0,\eta,\eta'\}$ decays at high energy, $\Delta\theta_{\gamma\gamma}\sim 2m_P/E_P$, increases the probability that both photons traverse mechanical gaps or non-instrumented apertures.  
In these configurations, the net transverse momentum carried by the undetected neutral system is typically negligible, 
recovering the elastic-balance pattern $|\mathbf p_{T,e'}|\simeq|\mathbf p_{T,p'}|$ despite hadronic emission.  
We mitigate these reducible backgrounds by imposing a global activity veto (including zero-degree and very-forward detectors), a narrow rejection band around $|\mathbf p_{T,e'}|\approx|\mathbf p_{T,p'}|$, and stability checks versus azimuth and pseudorapidity that expose crack-induced clustering.  
Events violating any of these criteria are rejected.

\emph{(3) Reconstruction tails and calibration biases.}  
Possible background sources of this type include: small angular biases at very forward electrons alter $Q^2=|t_e|$ and the transverse-momentum correlation, allowing elastic radiative events to evade elastic-balance rejections or, conversely, pushing signal-like events into vetoed regions; and cluster merging of collimated $\gamma\gamma$ from $\pi^0$ decays can also defeat photon identification and vetoes.  
It may be possible to control these effects by enforcing per-object pointing to the primary vertex, redundant timing for $e$ and $p$ consistent with the event time, cross-calibrated energy and angle scales (e.g., track-calorimeter consistency for $e$), and kinematic self-consistency tests requiring $M_X^2$ to be compatible with the measured $(\xi,t_p)$ within resolution. For many final states, it is reasonable to expect near 100\% vetoing efficiency, assuming they are within the $\eta < 4.0$ acceptance of ePIC. For final states with muons, experiments rely on minimum-ionizing signals in the hadronic calorimeters. Given the depth of these HCALs in ePIC (all $>$ 4 hadronic interaction lengths), the muons will interact with the HCAL volumes, but a full analysis of muon detection with appropriate thresholds would be better-suited to understand the full efficiency. In general, tightening the meson pseudorapidity cut to $\eta < 3.7-3.8$ to fully ensure that the daughters fall within the acceptance, and to ensure charged final states have both track and calorimeter information, would enhance the vetoing efficiency by reducing losses near fiducial acceptance boundaries. 

\subsection{Comment on reconstructing invariant mass}

An additional handle for background rejection for all three categories discussed above could, in principle, be obtained from reconstructing the invariant mass of the missing state $m_{\text{inv},X}(p_{e'},p_{p'})$. 
Namely, when targeting the invisible decays of a specific meson $P$, it is possible to impose a requirement $|m_{\text{inv},X}-m_{P}| < \delta$ to suppress contributions from other resonances and their visible decays. 
In practice, however, the missing mass resolution is highly sensitive to beam imperfections and to the kinematic smearing of $p_{e'}$ and $p_{p'}$, so even modest distortions can broaden the $m_{\text{inv},X}$ distribution to $\cO(100\%)$, making this handle very challenging to exploit. In addition, the $m_{\text{inv},X}$ reconstruction benefits only marginally from decreasing beam energy down to the minimal possible $E_{p} = 41\text{ GeV}$ and $E_{e} = 5\text{ GeV}$, since the intrinsic difficulty of reconstructing missing mass with smeared collider kinematics persists even at lower energies.

\section{EIC sensitivity estimation}
\label{app:event-selection}

In this section, we describe our calculation of the EIC sensitivity. 
We briefly discuss the impact of the selection criteria on the signal yield, as well as comment on the complications arising if considering the EIC setups with high-energy beam configuration. 
A more detailed discussion, including the investigation of the impact of various selection criteria on the signal yield, may be found in Ref.~\cite{Balkin:2026whv}.

\begin{table*}
\centering
\begin{tabular}{|c|c|c|c|c|c|}
\hline
$M$
& $\mathrm{BR}(P\to \mathrm{inv})_{\rm SM}$
& $\mathrm{BR}(P\to \mathrm{inv})_{\rm current}$
& Selection
& $\sigma_{\text{tot}}$ [fb]
& $\sigma_{\text{sel}}$ [fb]
\\
\hline\hline
$\pi^0$
& $4\times10^{-23}$~\cite{Gao:2018seg}
& $4.4\times10^{-9}$~\cite{NA62:2020pwi}
& \makecell{Baseline \\ Baseline$_{\text{FBD}}$ \\ Optimal}
& \makecell{$7.6\times10^{8}$ \\ $7.6\times10^{8}$ \\ $7.1\times10^{8}$}
& \makecell{$1.7\times10^{6}$ \\ $7.7\times10^{6}$ \\ $1.9\times10^{8}$}
\\
\hline
$\eta$
& $1\times10^{-18}$~\cite{Gao:2018seg}
& $1.1\times10^{-4}$~\cite{NA64:2024mah}
& \makecell{Baseline \\ Baseline$_{\text{FBD}}$ \\ Optimal}
& \makecell{$1.1\times10^{8}$ \\ $1.1\times10^{8}$ \\ $1.1\times10^{8}$}
& \makecell{$3.4\times10^{5}$ \\ $1.6\times10^{6}$ \\ $3.1\times10^{7}$}
\\
\hline
$\eta'$
& $5\times10^{-18}$~\cite{Gao:2018seg}
& $2.1\times10^{-4}$~\cite{NA64:2024mah}
& \makecell{Baseline \\ Baseline$_{\text{FBD}}$ \\ Optimal}
& \makecell{$3.0\times10^{7}$ \\ $3.0\times10^{7}$ \\ $2.8\times10^{7}$}
& \makecell{$1.2\times10^{5}$ \\ $5.0\times10^{5}$ \\ $8.2\times10^{6}$}
\\
\hline
$\rho^{0}$
& $2.4\times10^{-13}$~\cite{Schuster:2021mlr}
& --
& \makecell{Baseline \\ Baseline$_{\text{FBD}}$ \\ Optimal}
& \makecell{$6.3\times10^{9}$ \\ $6.3\times10^{9}$ \\ $5.4\times10^{9}$}
& \makecell{$1.6\times10^{7}$ \\ $3.5\times10^{7}$ \\ $6.8\times10^{8}$}
\\
\hline
$\omega$
& $2.8\times10^{-13}$~\cite{Schuster:2021mlr}
& $7.3\times10^{-5}$~\cite{ParticleDataGroup:2024cfk}
& \makecell{Baseline \\ Baseline$_{\text{FBD}}$ \\ Optimal}
& \makecell{$5.9\times10^{8}$ \\ $5.9\times10^{8}$ \\ $5.2\times10^{8}$}
& \makecell{$3.1\times10^{6}$ \\ $6.7\times10^{6}$ \\ $8.1\times10^{7}$}
\\
\hline
$\phi$
& $1.7\times10^{-11}$~\cite{Schuster:2021mlr}
& $1.7\times10^{-4}$~\cite{ParticleDataGroup:2024cfk}
& \makecell{Baseline \\ Baseline$_{\text{FBD}}$ \\ Optimal}
& \makecell{$1.5\times10^{8}$ \\ $1.5\times10^{8}$ \\ $1.2\times10^{8}$}
& \makecell{$8.4\times10^{5}$ \\ $1.6\times10^{6}$ \\ $8.5\times10^{6}$}
\\
\hline
\end{tabular}
\caption{Mesons $P=\pi^0,\eta,\eta',\rho^{0},\omega,\phi$, the branching ratios of their invisible decays as predicted in the SM, the current experimental bounds on their invisible decays, and the production cross sections before ($\sigma_{\text{tot}}$) and after ($\sigma_{\text{sel}}$) imposing the selection. We report the results for the Baseline and Optimal selection criteria from Tbl.~\ref{tab:selection} of the main text. Here, Baseline and Optimal denote the detector setup from Tbl.~\ref{tab:selection}, while Baseline$_{\text{FBD}}$ assumes the addition of the upgraded FBD detector, following the discussion in the main text.}
\label{tab:mesons_detailed}
\end{table*}

\begin{figure}[t!]
	\centering
    {\includegraphics[width=0.96\linewidth]{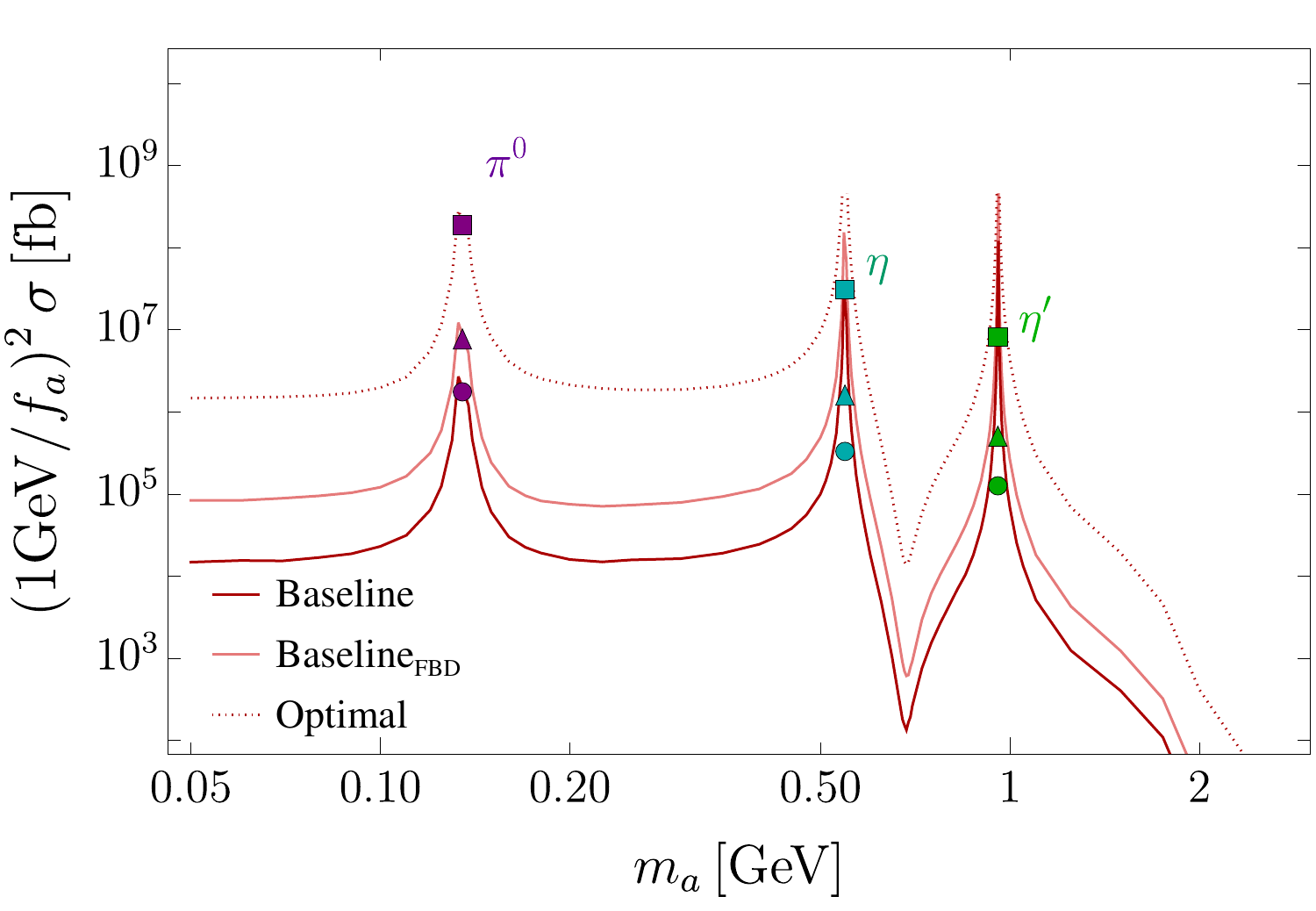}}
	\caption{
    The production cross section of the ALPs in the model of Eq.~\eqref{eq:Lagr_alp} with the reference choice $f_a=1\,\GeV$ with the two setups from Tbl.~\ref{tab:selection}: 
    Baseline (solid, dark), Baseline$_{\text{FBD}}$ (solid, pale), and Optimal (dashed). 
    We also show the $\pi^0$ and $\eta^{(\prime)}$ production cross sections 
    for the Baseline (circle), Baseline$_{\text{FBD}}$ (triangle), and Optimal (square) selections.}	
	\label{fig:xs}
\end{figure}

In Tbl.~\ref{tab:mesons_detailed}, we report the cross-section for each meson under Baseline and Optimal selection criteria from Tbl.~\ref{tab:selection}, before ($\sigma_{\text{tot}}$) and after ($\sigma_{\text{sel}}$) imposing selection criteria. We consider the pseudoscalar $\pi^{0},\eta,\eta'$ and vector $\rho^{0},\omega,\phi$ mesons. Finally, in Fig.~\ref{fig:xs}, we also report the ALP post-selection production cross-section $\sigma_{\rm sel}$. The calculation is based on Ref.~\cite{Balkin:2026whv}, which, following the description of Refs.~\cite{Aloni:2018vki,Ovchynnikov:2025gpx,Balkin:2025enj}, constructing the ALP production amplitude from the amplitudes of the pseudoscalar mesons $\pi^{0},\eta,\eta'$.

The selection efficiency (the ratio of $\sigma_{\rm sel}$ to the total production cross-section $\sigma_{\rm tot}$) is
\begin{align}
    \epsilon 
    \equiv 
    \frac{\sigma_{\text{sel}}}{\sigma_{\text{tot}}} 
    \simeq 
    \begin{cases} 
    0.002\text{--}0.006\,, \quad \text{Baseline} \\ 0.008\text{--}0.03, \quad \text{Baseline}_{\rm FBD}, \\  
     1/4\text{--}1/3 \,, \quad\quad\quad\quad\ \, \text{Optimal}
    \end{cases}\, .
\end{align}
Using $\sigma_{\rm sel}$, we estimate the EIC sensitivity:
\begin{align}
    &\sum_{X = a,\pi^{0},\eta,\eta',\rho^{0},\omega,\phi} \cL\times \sigma_{\rm sel} > \begin{cases}
        3\,, \;\;\;\;\;\; & N_{\text{bg}}=0
        \\
        2\sqrt{N_{\text{bg}}}\,, & \text{otherwise}
    \end{cases}\,.
\end{align}
Now, let us comment on the complexities of utilizing the MPE search for the EIC setups with high total $ep$ energy. Increasing the energy shifts the kinematics of the $X$ state to the partially instrumented domain $\eta_{X}>4$, rendering the $\eta_{X}<4$ cut very harmful to the signal event yield.
This problem may be mitigated by adding the partially instrumented $4.5<\eta_X<6$ domain to the event selection. 
The price for this is a significantly increased vetoing inefficiency. 
A more detailed investigation of the inefficiencies for the $4.5<\eta<6$ window requires full simulation. 
However, for illustrative purposes, we considered the configuration with the highest possible energy, $E_{p} = 275\,\GeV$, $E_{e} = 18\,\GeV$, and repeated the same fast simulation we conducted for the $(100,10)\,\GeV$ configuration. 
Compared to the energy configuration of the Baseline setup, we found that the vetoing inefficiency increases to $2\%$ for $\rho^{0}\to \pi^{+}\pi^{-}$, $4\%$ for $\phi\to K\bar{K}$, 
$0.2\%$ for $\pi^{0}\to e^{+}e^{-}$, and 5\% for $\pi^{0}\to\gamma\gamma$.

The inefficiency may be reduced by increasing the instrumentation of the $\eta_{X}>4$ region. 
For the ePIC detector, this is very complicated due to the lack of space for new detectors, but it may be possible for the possible second detector.

\section{More on invisible signatures}
\label{app:meson-to-invis}

In this section, we compare signatures with invisible decays arising in the models with hadronically coupled ALPs.

The generic Lagrangian reads
\begin{align}
    \label{eq:ALP-model-SuM}
	\mathcal L_a 
    \supset 
    &-\frac{c_g\alpha_s}{8\pi } \frac{ a }{f_a} G^{\mu\nu} \tilde{G}_{ \mu \nu} 
	+\frac{\partial_\mu a}{f_a} \bar{q} \, c_q\, \gamma^\mu\gamma_5 q \nonumber\\
    &-\bar{q} M_qe^{2i\gamma^5\beta_q \frac{a}{f_a}} q+ ig_{a\chi}a\bar \chi \gamma^{5}\chi \,.
\end{align}
It includes not only gluonic coupling, but also couplings to quarks in the axial-vector current, $c_{q}$, and in the quark mass term, $\beta_{q}$.
Additional couplings that do not contribute to the relevant processes have been omitted.
The model presented in Eq.~\eqref{eq:Lagr_alp} is equivalent to taking $c_g=-2$ and $c_q=\beta_q=0$.

We define the following combinations of the ALP couplings:
\begin{align}
    \label{eq:ALP-invariant-couplings}
	\bar{c}_q\equiv c_q+\beta_q\,,\qquad 
    \bar{c}_g\equiv c_g+\langle \beta_q\rangle  \,,
\end{align}
where $\langle...\rangle=2Tr(...)$, with the trace being performed in flavor space.
These combinations have the property of being invariant under an axial rotation of the quark, and thus measurable quantities may depend on them~\cite{Balkin:2025enj}. 

The two contributions to the invisible events in this model are: 
(i) decays of \emph{on-shell ALPs}, $a\to \inv$, 
and (ii) decays of pseudoscalar mesons $P$ via \emph{off-shell ALPs}, $P\to a^{*}\to \inv$.

The calculation of the ALP production cross-section, which determines the sensitivity to the on-shell contribution, is discussed in detail in Ref.~\cite{Balkin:2026whv}; 
we refer to this work and do not provide details here. 
The branching ratio of invisible ALP decays reads
\begin{align}
    \label{eq:ALP-inv}
    \BR(a\to\inv) 
    = \frac{\Gamma(a\to\chi\bar{\chi})}{\Gamma(a\to\chi\bar{\chi})+\Gamma(a\to{\rm vis})}\,,
\end{align}
where $\Gamma(a\to\chi\bar{\chi}) = g_{a\chi}^{2}\sqrt{m_{a}^{2}-4m_{\chi}^{2}}/8\pi$ is the invisible width and $\Gamma(a\to{\rm vis})\propto f_{a}^{-2}$ is the decay width into visible states originating from the coupling to hadrons (see Refs.~\cite{Ovchynnikov:2025gpx,Balkin:2025enj}). 
Parametrically, for allowed couplings $f_{a}$, $\BR(a\to\inv)$ is very close to unity for $g_{a\chi}\gtrsim 10^{-3}$ and even smaller, depending on the ALP mass.
 
Now, let us proceed to the off-shell contribution. 
Below, we calculate the branching ratios of the pseudoscalar mesons $\pi^{0}$ and $\eta^{(\prime)}$ into the invisible state $\chi\bar{\chi}$ and then discuss the range of ALP masses and couplings probed by the two contributions.

Decays of pseudoscalar mesons via off-shell ALPs may be interpreted in terms of mixing between the ALPs and $P$.
The relation between the interaction states and the mass states (labeled as 'phys') is given by
\begin{align}
    a 
    = 
    a_{\rm phys}+\frac{f_\pi}{f_a}\sum_{P=\pi,\eta,\eta'} g_{aP} P_{\text{phys}} \, ,
\end{align}
where $f_\pi=92\,\MeV$. 
The coefficients $g_{aP}$ are calculated using the approach of Refs.~\cite{Aloni:2018vki,Ovchynnikov:2025gpx,Balkin:2025enj}:
\begin{align}
    \label{eq:ALP-pi-generic}
    g_{a\pi} 
    =& 
    \frac{m_\pi^2}{m_a^2-m_\pi^2}
    \Bigg\{
    (\bar{c}_u-\bar{c}_d)
    +\frac{\delta_I}{4}\left[-2\bar{c}_g+\frac{3m_\pi^2}{(m_\eta^2-m_\pi^2)}\left(\bar{c}_u+\bar{c}_d-\frac{2}{3}\bar{c}_s\right)\right]
    \Bigg\}\, ,      
    \\
    \label{eq:ALP-eta-generic}
    g_{a\eta} 
    =&
    \frac{2}{\sqrt{6}}\frac{1}{m_a^2-m_\eta^2}
    \Bigg\{
    m_\eta^2\left(-\frac{\bar{c}_g}{2}+\bar{c}_u+\bar{c}_d-\bar{c}_s\right)
    +m_\pi^2\frac{\bar{c}_g}{2}
    +\delta_I\frac{m_\pi^2m_\eta^2}{m_\eta^2-m_\pi^2}\left(\bar{c}_d-\bar{c}_u\right) 
    \Bigg\}\, ,    
    \\
    \label{eq:ALP-etapr-generic}
    g_{a\eta'} 
    =& 
    \frac{4}{\sqrt{3}}\frac{1}{m_a^2-m_{\eta'}^2}
    \Bigg\{
    \frac{m_\eta^2}{2}\left(-\bar{c}_g+2\bar{c}_u+2\bar{c}_d+4\bar{c}_s\right)
    +\frac{m_\pi^2}{4}\left(2\bar{c}_g-3\bar{c}_u-3\bar{c}_d-6\bar{c}_s\right)
    +\frac{\delta_I}{4}\frac{m_\pi^2m_{\eta'}^2}{m_{\eta'}^2-m_\pi^2}\left(\bar{c}_d-\bar{c}_u\right)\Bigg\} \, , 
\end{align}
where $\delta_I\equiv(m_d-m_u)/(m_d+m_u)$ is the isospin-breaking parameter.

For the gluonic ALP model (Eq.~\eqref{eq:Lagr_alp}), these are:
\begin{align}
    g_{a\pi} 
    =& 
     \delta_I
    \frac{m_\pi^2}{m_a^2-m_\pi^2} \, , 
       \label{eq:ALP-pi} 
    \\
    g_{a\eta} 
    =&
    \frac{2}{\sqrt{6}}\frac{m_\eta^2-m_\pi^2}{m_a^2-m_\eta^2} \, ,    
    \label{eq:ALP-eta}
    \\
    g_{a\eta'} 
    =& 
    \frac{4}{\sqrt{3}}\frac{m_{\eta}^2-m_\pi^2}{m_a^2-m_{\eta'}^2} \, .     
    \label{eq:ALP-etapr}
\end{align}
The resulting branching ratio of the process $P\to\bar{\chi}\chi$ via off-shell ALP is 
\begin{align}
    \label{eq:off-shell-width}
    \BR(P\rightarrow\bar{\chi}\chi)
    = 
    \frac{m_P}{8\pi \Gamma_{P}}
    \left(\frac{f_\pi}{f_a}g_{aP}g_{a\chi}\right)^2
    \sqrt{1-\frac{4m_\chi^2}{m_P^2}}\, ,
\end{align}
where $\Gamma_{P}$ is the total decay width of $P$. 

As a consequence, we can recast the projected EIC sensitivity to invisible meson decays, 
$[\BR(P\to\inv)]_{\rm EIC}$, 
under the assumed background scenario, into a corresponding projected reach in the coupling combination
$g_{a\chi}/f_a$:
\begin{align}
    \label{eq:PinvMaster}
    \frac{g_{a\chi} }{f_a}
    >
    \frac{1}{f_\pi g_{aP}}
    \sqrt{8\pi\Gamma_P\frac{[\BR(P\rightarrow\text{inv})]_{\text{EIC}}}{\sqrt{m_P^2-4m_\chi^2}}} \, .
\end{align}

For the case of the gluonic ALPs with $c_{g} = -2$, $\beta_{q}=c_{q} = 0$, using Eqs.~\eqref{eq:ALP-pi}--\eqref{eq:PinvMaster} for $m_\chi=0$ and $m_a\gg m_P$, we can derive the asymptotic form of the EIC reach:
\begin{align}
    m_a\sqrt{\frac{f_a/g_{a\chi}}{ 10^2\,\GeV}}
    >&
    10\left(\frac{2\times 10^{-9}}{[\BR(\pi^0\rightarrow\text{inv})]_{\text{EIC}}}
    \right)^{1/4}\GeV \, ,\\
    m_a\sqrt{\frac{f_a/g_{a\chi}}{ 10^2\,\GeV}}
    >&
    17\left(\frac{1\times 10^{-8}}{[\BR(\eta\rightarrow\text{inv})]_{\text{EIC}}}
    \right)^{1/4}\GeV \, ,\\
    m_a\sqrt{\frac{f_a/g_{a\chi}}{ 10^2\,\GeV}}
    >&
    6\left(\frac{8\times 10^{-8}}{[\BR(\eta^\prime\rightarrow\text{inv})]_{\text{EIC}}}
    \right)^{1/4}\GeV \, .
\end{align}
We learn that invisible meson decays at the EIC will be sensitive to ALPs with $m_a\sim\cO(10\,\GeV)$  for $f_a/g_{a\chi}\sim100\,\GeV$. 

\begin{figure}[t!]
    \centering
    \includegraphics[width=0.48\textwidth]{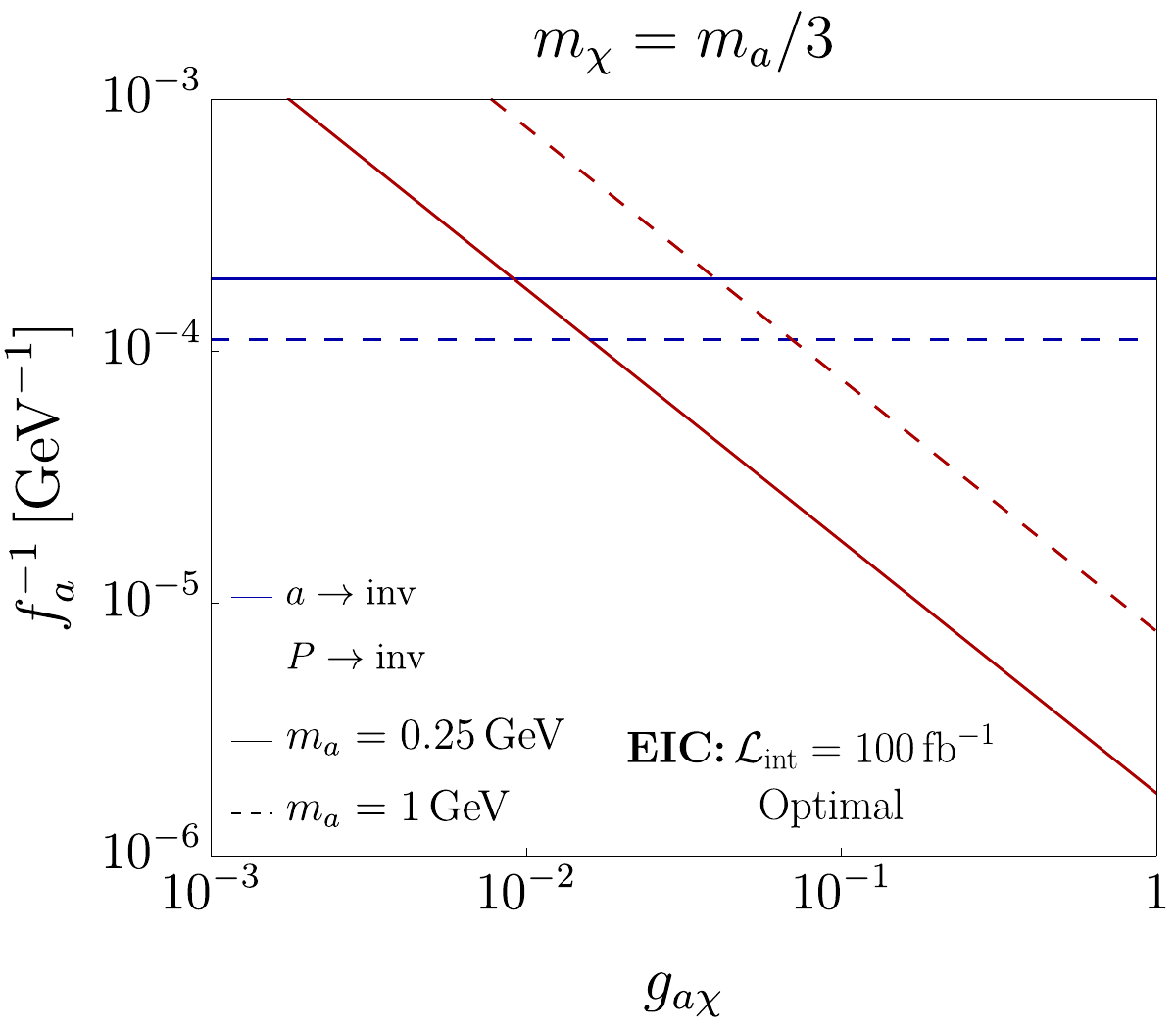}
	\caption{Complementarity between two contributions to the missing-proton-energy events at the EIC: decays of pseudoscalar mesons (red) and of ALPs (blue), for two ALP masses $m_{a} = 0.25\text{ GeV}$ (solid) and $1\text{ GeV}$ (dashed), in the plane $\{g_{a\chi}, f_{a}^{-1}\}$. The curves show the lower bound of the sensitivity corresponding to the Baseline selection, under the assumption of zero background.}
	\label{fig:complementarity-signatures}
\end{figure}

Next, we discuss the complementarity of the on-shell and off-shell ALP contributions to the MPE events. The yield of events with on-shell ALPs scales with $\BR(a\to\inv)/f_{a}^{2}$, whereas the number of events with off-shell ALPs behaves as $g_{a\chi}^{2}/f_{a}^{2}$. As a result, they prefer different domains of the parameter space. In particular, given the behavior of $\BR(a\to \inv)$, the on-shell contribution is the main one for small $g_{a\chi}\ll 1$. On the other hand, the off-shell events dominate for larger $g_{a\chi}$.
We demonstrate the EIC sensitivity to these two cases in the plane $g_{a\chi}-f_{a}^{-1}$ in Fig.~\ref{fig:complementarity-signatures}.

\section{Interpreting current bounds from invisible meson decays}
\label{app:na64h}

We estimate the bounds on the gluonic ALP model of Eq.~\eqref{eq:ALP-model-SuM} coming from existing measurements of invisible meson decays.
The NA62 collaboration reported bounds on invisible decays of the $\pi^{0}$ meson in Ref.~\cite{NA62:2020pwi} and the process $K^{+}\to \pi^{+}+\text{inv}$ in Ref.~\cite{NA62:2025upx}. 
In both cases, the missing-mass technique has been adopted, so these are independent analyses disentangling the off-shell and on-shell ALP decays.
The resulting upper limit on the allowed branching ratio reads
\begin{align}
    &[\BR(\pi^{0}\to \rm{inv})]_{\text{NA62}}
    <
    4.4\times 10^{-9}\,,\\ 
    &[\BR(K^{+}\to \pi^{+}+\rm{inv})]_{\text{NA62}}
    \lesssim 
    2\times 10^{-11}\, ,
\end{align}
where the second bound is mass-dependent. 

Within the model of Eq.~\eqref{eq:ALP-model-SuM}, $\BR(\pi^{0}\to \bar{\chi}\chi)$ is given by Eq.~\eqref{eq:off-shell-width}. 
The $K\to \pi \chi\bar{\chi}$ rates can be written as
\begin{align}
    \BR(K\to \pi \chi\bar{\chi}) 
    = \BR(K^{+}\to \pi^{+}a) \BR(a\to \chi\bar{\chi})\,,
\end{align}
where $\BR(K^{+}\to \pi^{+}a)$ is given in Ref.~\cite{Bauer:2021wjo} and $\BR(a\to \chi\bar{\chi})$ is Eq.~\eqref{eq:ALP-inv}.

Next, we reinterpret the NA64h result, where a 50\,GeV pion beam hits a fixed target. NA64h reported bounds on the invisible decay of particles produced in the process~\cite{NA64:2024mah}
\begin{align}
    \label{eq:NA64-process}
    \pi^{-}+p\to n+\text{inv} \, .
\end{align}
The missing invariant mass was not reconstructed; therefore, the bound is imposed on the sum $\sum_{X = \pi^{0},\eta,\dots}N_{X}\cdot \BR(X\to \text{inv})$. 
Under the assumption that solely $\eta\,(\eta')$ meson dominates the invisible event yield, the bounds read
\begin{align}
    \label{eq:NA64h-eta}
    &[\BR(\eta\to \text{inv})]_{\rm NA64h}=1.1\times 10^{-4}\, , \\ 
    \label{eq:NA64h-etapr}
    &[\BR(\eta'\to \text{inv})]_{\rm NA64h}=2.1\times 10^{-4}\, .  
\end{align}

In the context of our ALP model, Eq.~\eqref{eq:ALP-model-SuM}, the bounds from off-shell ALP production can be read as
\begin{align}
    \label{eq:bound-meson-NA64}
    \sum_{P = \eta,\eta'} 
    \frac{\BR(P\to \bar{\chi}\chi)}{[\BR(P\to \text{inv})]_{\text{NA64h}}} < 1\,.
\end{align}
For the case when on-shell ALP decays dominate, we would need to know the event rate after imposing the selection, which is obviously ALP-mass-dependent. 
However, we estimate it by using two assumptions. 
\emph{First}, we assume that the matrix element of the process~\eqref{eq:NA64-process} with the production of $P = \pi^{0},\eta,\eta',a$ is dominated by the $t$-exchange of vector mesons $V = \rho, \phi, \omega$ and their excitations. 
The contribution from individual $V$-exchange scales as $G_{V}(t)\times g_{\pi VP}\times g_{Vpn}$, with $G_{V}$ being reggeized propagator, $g_{\pi VP}$ the $\pi^{-}V^{+}P$ constant, and $g_{Vpn}$ the $V^{-}pn$ constant (see Ref.~\cite{Balkin:2026whv}). 
Within the universal Regge-parameterization of the cross-section from~\cite{Gninenko:2023rbf} utilized in the NA64h analysis, it allows us to assume 
\begin{align}
    \sigma_{\text{tot}}(\pi+p\to P+n)
    \propto 
    \left|\sum_{V}g_{\pi VP}g_{Vpn}\right|^{2} \, ,
    \label{eq:sigma-scaling-NA64}
\end{align}
for the total cross-section before imposing analysis-driven selection. 
Second, Ref.~\cite{NA64:2024mah} mentions that the characteristic signal shape for $\eta,\eta'$ events would be very similar. 
Therefore, for a good approximation, the selections are independent of $P$ (or $m_a$).

Following the above arguments, we estimate the NA64h bound on invisibly decaying ALPs in the following way:
\begin{align}
    &\frac{N_{a\to \inv}}{[N_{\eta^{'}\to \inv}]_{\text{NA64h}}} 
    \nonumber\\ 
    &= \frac{\sigma_{\text{tot}}(\pi+p\to a+n)}{\sigma_{\text{tot}}(\pi+p\to \eta'+n)}\frac{\BR(a\to\inv)}{[\BR(\eta'\to \text{inv})]_{\rm NA64h}} 
    \lesssim 1  \, .
\end{align}
where $N_{a\to \inv}$ is the number of invisible ALP decays within the NA64h selection, while $[N_{\eta^{'}\to \inv}]_{\text{NA64h}}$ is the maximally allowed number of invisibly decaying $\eta'$, which is given by Eq.~\eqref{eq:NA64h-etapr}. The ratio of the cross-sections is given by Eq.~\eqref{eq:sigma-scaling-NA64}, where the coupling $g_{\pi Va}\propto f_{a}^{-1}$.
We have found that the bound does not exceed the $B\to K+\inv$ and $K\to \pi+\inv$ bounds except for the narrow vicinities of the $\eta,\eta'$ masses; therefore, we neglect it. 
Accurate calculations of this bound and the NA64 sensitivity are left for future work.

\clearpage

\end{document}